\documentclass[fleqn]{natureprintstyle}
\usepackage[T1]{fontenc}

\usepackage{authblk}

\usepackage{amsmath}
\usepackage{amsfonts}
\usepackage{amssymb}
\usepackage{graphicx}
\usepackage{subcaption}
\usepackage{csquotes}

\usepackage{hyperref}
\usepackage{float}

\title{The diffuse $\gamma$-ray background is dominated by star-forming galaxies}

\author[1,*]{Matt A. Roth}
\author[1,2]{Mark R. Krumholz}
\author[1]{Roland M. Crocker}
\author[3]{Silvia Celli}
\affil[1]{Research School of Astronomy and Astrophysics, The Australian National University, Canberra, Australia}
\affil[2]{ARC Centre of Excellence for All-Sky Astrophysics in Three Dimensions (ASTRO-3D), Canberra, Australia}
\affil[3]{Dipartimento di Fisica dell'Universit\`a La Sapienza and INFN, P.~le Aldo Moro 2, 00185 Rome, Italy}
\affil[*]{matt.roth@anu.edu.au}

\begin{document}

\flushbottom
\maketitle

\section*{Published in \href{https://www.nature.com/articles/s41586-021-03802-x}{Nature DOI 10.1038/s41586-021-03802-x}}

\thispagestyle{empty}

\begin{abstract}

\textbf{The \textit{Fermi} Gamma-ray Space Telescope has revealed a diffuse $\gamma$-ray background at energies from 0.1 GeV to 1 TeV, which can be separated into Galactic emission and an isotropic, extragalactic component \textsuperscript{1}. Previous efforts to understand the latter have been hampered by the lack of physical models capable of predicting the $\gamma$-ray emission produced by the many candidate sources, primarily active galactic nuclei \textsuperscript{2-5} and star-forming galaxies \textsuperscript{6-10}, leaving their contributions poorly constrained. Here we present a calculation of the contribution of star-forming galaxies to the $\gamma$-ray background that does not rely on empirical scalings, and is instead based on a physical model for the $\gamma$-ray emission produced when cosmic rays accelerated in supernova remnants interact with the interstellar medium \textsuperscript{11}. After validating the model against local observations, we apply it to the observed cosmological star-forming galaxy population and recover an excellent match to both the total intensity and the spectral slope of the $\gamma$-ray background, demonstrating that star-forming galaxies alone can explain the full diffuse, isotropic $\gamma$-ray background.}

\end{abstract}

Many candidate sources have been proposed for the origin of the diffuse, isotropic $\gamma$-ray background. These include active galactic nuclei (AGN) (particularly blazars \textsuperscript{2-5}), millisecond pulsars \textsuperscript{2}, star-forming galaxies (SFGs) \textsuperscript{6-10}, and dark matter annihilation \textsuperscript{12}. Previous estimates of their contributions have relied on a highly-uncertain process of empirically scaling the emission from a small sample of local, resolved sources by their estimated cosmological abundances, whereas our approach in this paper is instead to calculate the emission from SFGs directly using a physical model.
The cosmic rays (CRs) responsible for $\gamma$-ray emission in SFGs (including the Milky Way) are produced by diffusive acceleration at supernova remnant shocks \textsuperscript{13}. This process transfers $\sim$10\% of the supernova mechanical energy to relativistic ions, yielding on average $\sim 10^{50}$ erg in CR ions per supernova \textsuperscript{14,15}, with another $\sim$2\% ($\sim 2\times 10^{49}$ erg) deposited in CR electrons \textsuperscript{16}. The resulting CRs follow a power law distribution in particle momentum $p$ of the form $dn/dp \propto p^{-q}$ \textsuperscript{17,18}; observations of individual supernova remnants, analytical models, and numerical simulations all indicate that the index $q$ is in the range $q \approx 2.0 - 2.6$, with a mean value of $q\approx 2.2-2.3$ \textsuperscript{19,20}. Some of the CR ions collide inelastically with interstellar medium (ISM) nuclei, producing roughly equal numbers of $\pi^0$, $\pi^+$ and $\pi^-$ mesons that rapidly decay via the channels $\pi^0 \to 2\gamma$, $\pi^- \to \mu^- + \bar{\nu}_\mu$, and $\pi^+ \to \mu^+ + \nu_\mu$. The decay of $\pi^0$ particles is responsible for most of the observed Galactic $\gamma$-ray foreground, which displays a characteristic spectrum that rises sharply from $\sim 0.1-1$ GeV as a result of the $135$ MeV rest mass of the $\pi^0$ particle.

As this discussion suggests, a SFG's diffuse $\gamma$-ray emission depends primarily on three factors: its total star formation rate (which determines its supernova rate and thus the rate at which CRs are injected), the distribution of $\gamma$-ray energies produced when individual CRs collide with ISM nuclei (which depends on the parent CR energy $E$), and the fraction of CRs (again as a function of $E$) that undergo inelastic collisions before escaping the galaxy. The first two of these are relatively well-understood, but the third factor, known as the calorimetry fraction $f_{\rm cal}(E)$, is much less certain. It depends on the properties of the galaxy, and the lack of a model for this dependence has previously precluded direct calculation of SFG $\gamma$-ray emission. However, Ref.~\textsuperscript{11} recently introduced a model for $f_{\rm cal}(E)$, based on rates of CR diffusion determined by the balance between the CR streaming instability and ion-neutral damping. In the Methods we describe a new technique to use this model to compute $f_{\rm cal}(E)$, and thus the total $\gamma$-ray emission produced by CR ions in SFGs.

We supplement the $\gamma$-ray production rate from CR ions by adding the contribution from both primary CR electrons, directly injected by supernova remnants, and secondary CR leptons (electrons and positrons), produced in the $\pi^{\pm}$ decay chain; these become important at energies $\lesssim 1$ GeV. Our model for these particles includes energy losses due to ionisation, synchrotron emission, bremsstrahlung, and inverse Compton scattering. The model also includes the attenuation of $\gamma$-rays produced by both CR ions and leptons due to pair production in collisions with far-infrared photons inside the source galaxy and extragalactic background light photons outside the galaxy, which become important at energies $\gtrsim 100$ GeV. The radiation that is absorbed by the host galaxy and extragalactic photon fields is reprocessed to lower energies in a pair-production cascade, whereby the initial high energy pair inverse Compton scatters lower energy photons up to $\gamma$-ray energies and these, in turn, produce further pairs, and so on. Details of our calculation of all these processes are provided in the Methods.

We now have a model that predicts the $\gamma$-ray emission of a SFG. The next step in our analysis is to apply this model to a galaxy survey that samples the SFG population out to the epoch of peak cosmological star formation at $z \sim 2$. For this purpose, we make use of the Cosmic Assembly Near-infrared Deep Extragalactic Legacy Survey (CANDELS) \textsuperscript{21,22} in the GOODS-S field. We apply our model to the CANDELS sample as described in the Methods. To verify that our approach predicts reasonably accurate $\gamma$-ray spectra, we apply it to four local, resolved galaxies with measured $\gamma$-ray emission \textsuperscript{7,23,24},
chosen to span a wide range of gas and star formation surface densities: Arp 220, NGC 253, M31, and NGC 4945. The input data we use for these calculations are summarised in Extended Data Table 1, and we show the results of the computation in Figure 1, where the solid lines show the spectra derived using only stellar data (as we have for CANDELS) and, for comparison, the dotted lines show the results we obtain if we add directly-measured gas properties (available for these local galaxies). We see that the fits are slightly improved if we make direct use of gas data but, even for the stellar data only, our model reproduces the observed $\gamma$-ray spectra to better than a factor of 2 for all galaxies at energies $>1$ GeV, and within a factor of $\approx 1.5$ for the two more rapidly star-forming galaxies, which, as we show below, are more akin to the population that dominates the $\gamma$-ray background.

\begin{figure}
  \includegraphics[width=0.49\textwidth]{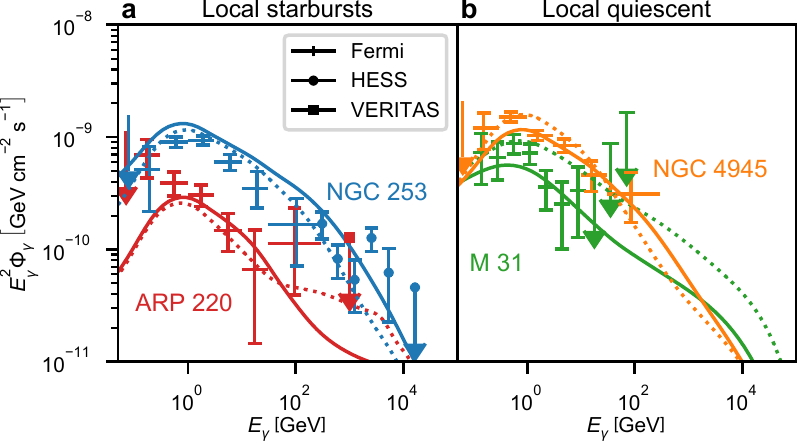}
  \caption{\textbf{The $\gamma$-ray spectra of nearby SFGs} Predicted (lines) and observed (points) spectra for a selection of nearby SFGs detected in $\gamma$-rays. The observations shown are taken from a combination of \textit{Fermi} LAT \textsuperscript{23}, HESS \textsuperscript{7}, and VERITAS \textsuperscript{24} where the horizontal bars show the energy bin and the vertical bars the 1 $\sigma$ uncertainty limit; in Panel \textbf{a} we show the local starburst galaxies Arp 220 and NGC 253, and Panel \textbf{b} we show the local quiescent galaxies M31 and NGC 4945. The solid lines show model predictions using only stellar data of the type we have available for the CANDELS sample, while the dotted lines shows results predicted if we supplement this with observed gas data. We list the full set of observed quantities used in computing these models in Extended Data Table 1.}
\end{figure}

Having verified that we can obtain accurate predictions of $\gamma$-ray spectra from stellar data alone, we carry out two additional validation steps. First, we examine the correlation between galaxies' far-infrared and total $\gamma$-ray luminosities, computed as described in the Methods. In Figure 2 we show the resulting distribution of galaxies in the $L_{\rm FIR}$ - $L_{\gamma}$ plane, along with a power law fit to the data (blue line), $L_\gamma/{\rm erg \; s^{-1}} = 10^{28.26 \pm 1.55} (L_{\rm FIR}/{\rm L_\odot})^{1.14 \pm 0.16}$. Our model prediction shows good agreement with the observed relation \textsuperscript{25}, and we note that both the model and the observed correlation differ noticeably from the calorimetric limit obtained by simply setting $f_{\rm cal}\left(E\right) = 1$ (red line in the Figure). Thus the agreement is non-trivial, and suggests that our model is correctly predicting the variation in galaxies' calorimetry fractions as a function of star formation rate.

\begin{figure}
  \includegraphics[width=0.49\textwidth]{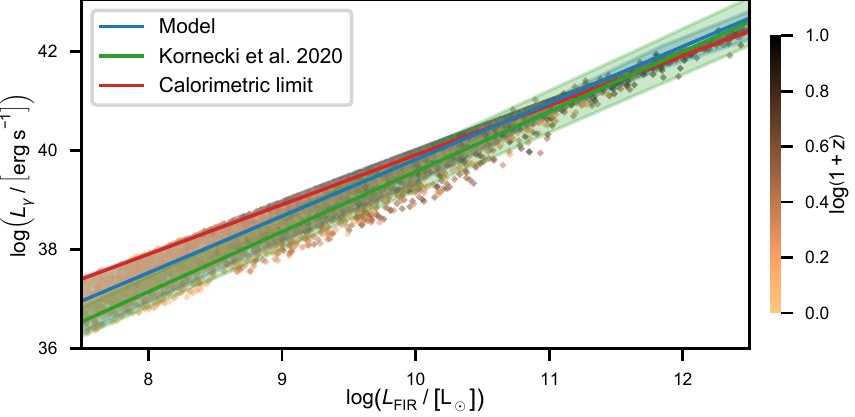}
  \caption{\textbf{The FIR-$\gamma$ correlation} The correlation between far-infrared ($8-1000$ $\mu$m) and $\gamma$-ray ($0.1-100$ GeV) luminosity for the CANDELS sample, derived using our model. Points show individual CANDELS galaxies, colour-coded by redshift $z$. The blue line is a power law fit to the CANDELS sample with the shaded band containing 90\% of data around the model fit. For comparison, the solid green line shows the empirical relation measured for 14 nearby, resolved SFGs \textsuperscript{25} with 2 $\sigma$ uncertainty in the shaded band. The red line is the calorimetric limit obtained by taking $f_{\rm cal} = 1$ at all energies in \autoref{eq:Lgamma}, as obtained by Ref.~\textsuperscript{7}.}
\end{figure}

Our second validation test is to compare our model with counts of \textit{resolved} SFGs observed by \textit{Fermi} LAT. Details of how we perform the comparison are given in the Methods. We show the results in Figure 3, which demonstrates that our model predicts SFG source counts consistent with observations, with the exception that we do not predict sources as bright as the Milky Way's two satellite galaxies, the Large and Small Magellanic Clouds. This is not surprising, since our comparison includes only field galaxies.

Our final step is to compute the contribution of SFGs to the diffuse, isotropic $\gamma$-ray background (see Methods for details). We present the results of this calculation in Figure 4, and provide a detailed analysis of the model uncertainties in the Supplementary Information and Extended Data Figure 1. Figure 4 shows that the expected contribution of SFGs to the diffuse isotropic $\gamma$-ray background fully reproduces both the intensity and the spectral shape of the observations from $\approx 0.2$ GeV to $\approx 1$ TeV. We emphasise that we obtain this agreement from our model with no free parameters: our only inputs are the CR injection spectral index ($q=2.2$), the energy per supernova ($10^{51}$ erg), and the fraction of supernova energy that goes into primary ions and electrons (10\% and 2\%, respectively) -- all quantities that are directly measured in the local Universe -- and the distribution of SFGs sampled by CANDELS. The key to the success of the model is the galaxy-by-galaxy calculation of the energy-dependent calorimetry fraction $f_{\rm cal}(E)$, which we demonstrate by also plotting the result (dotted line) we would obtain simply by setting $f_{\rm cal} = 1$ for all galaxies at all energies. This clearly both overestimates the intensity and yields a spectral slope that is flatter than observed.

We show the relative contributions to the background made by galaxies with differing star formation rates and redshifts in Extended Data Figure 2. The Figure shows that the background at lower energies is dominated by galaxies from just after cosmic noon ($z\sim 1-2$), while at higher energies, where attenuation by extragalactic background light has a larger effect, the dominant contribution shifts towards lower redshifts, so that at 1 TeV the background is dominated by $z\sim 0.1$ sources. At all energies, the dominant contribution comes from galaxies at the upper end of the star-forming main sequence, which have high but not extreme star formation rates for their redshift.

\begin{figure*}
  \centering
  \includegraphics[width=0.69\textwidth]{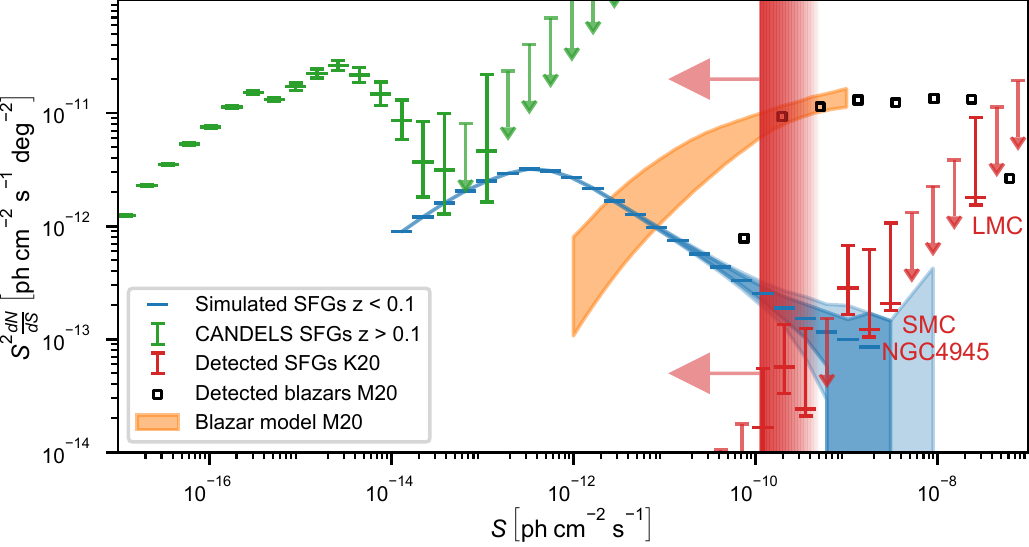}
  \caption{\textbf{The $\gamma$-ray source count distribution} Red points show the compensated distribution of SFG luminosities \textsuperscript{25} $S^2 (dN/dS)$ on the sky as a function of photon flux $S$ integrated from 1 - 100 GeV as seen by \textit{Fermi} LAT; error bars show 90\% confidence intervals or upper limits. The three brightest non-empty bins each contain only a single SFG, which we have labelled. Green points show model-predicted source counts for observed CANDELS galaxies at $z>0.1$ with the 90\% confidence limit, and blue points show Monte Carlo realisations of the $z<0.1$ SFG population, with the light and dark shaded bands indicating 68\% and 90\% confidence intervals. Black squares show \textit{Fermi}-detected blazars, and the orange band shows the blazar distribution model of Ref.~\textsuperscript{27} within the 1 $\sigma$ range. Finally, the red vertical band indicates the flux range over which \textit{Fermi} observations become incomplete; the left edge of this band is the 4FGL threshold for 98\% detection efficiency for sources with spectral index 2.3 \textsuperscript{27}.}
\end{figure*}

It is important to put our finding that SFGs dominate the diffuse, isotropic $\gamma$-ray background in the context of recent work, where a number of authors have argued that blazars and other AGN sources contribute substantially or even dominate the background. We provide a more detailed discussion in the Supplementary Information, but here note that we find that, while blazars dominate the \textit{resolved} component of the extragalactic $\gamma$-ray background, as shown in Figure 3, SFGs dominate the unresolved component. This finding is consistent with statistical analyses of angular fluctuations in the isotropic background and cross-correlations between it and galaxies and quasars, which strongly disfavour blazars as a dominant contributor \textsuperscript{2,8,26}. Indeed, a straightforward extrapolation of the number counts of observed blazars \textsuperscript{27}, illustrated by the orange band in Figure 3, also suggests that blazars do not dominate the unresolved background. Our finding that SFGs alone are able to reproduce the full background is also consistent with the conclusions of Refs.~\textsuperscript{9} and \textsuperscript{28} that, in the absence of either a physical model for the $\gamma$-ray emission of SFGs or a much larger sample of resolved galaxies, it is not possible to rule them out as a dominant contributor.

\begin{figure}[H]
  \centering
  \includegraphics[width=0.49\textwidth]{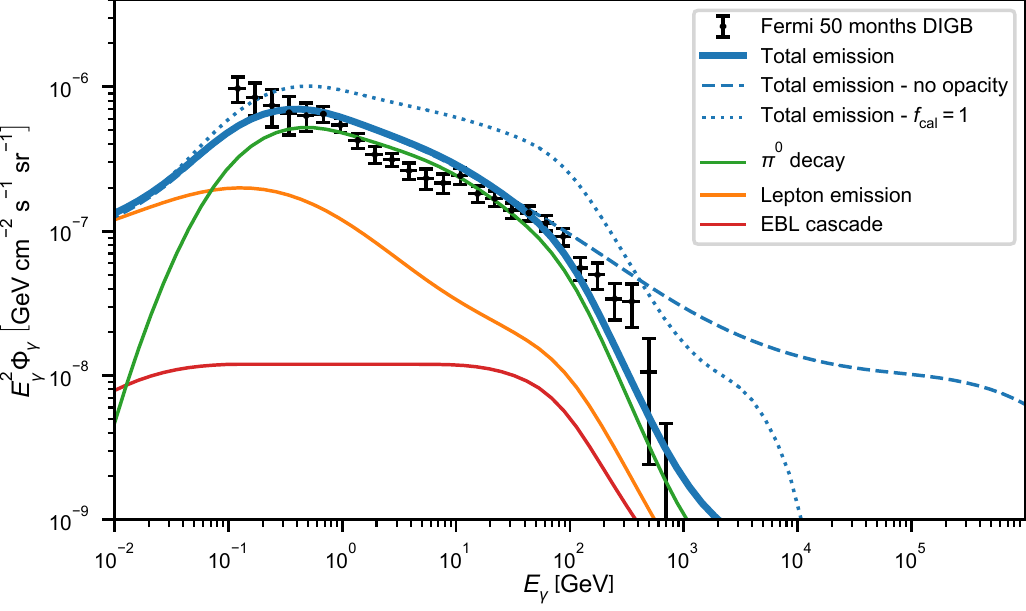}
  \caption{\textbf{The diffuse isotropic $\gamma$-ray background} Black data points show \textit{Fermi} 50-month observations \textsuperscript{1} where the horizontal bars show the energy bin and the vertical bars the 1 $\sigma$ uncertainty limit; the thick blue line shows our prediction for the total background due to SFGs; thin solid lines in other colours show the fractional contribution to this total from $\pi^0$ decay (CR ions; green line), leptonic emission processes (CR electrons and positrons; orange line), and $\gamma\gamma$ scattering and the resulting pair production cascade (red line). Broken thin blue lines show the predicted background if we turn off parts of our model: the blue dotted line shows the spectrum we would obtain if we set $f_{\rm cal} = 1$ for all galaxies at all energies, while the blue dashed line shows the background we would see in the absence of $\gamma\gamma$ opacity.}
\end{figure}

We conclude by pointing out that the methodology we have introduced can also be applied to predict luminosity functions and background contributions from SFGs at other wavelengths and in other messengers driven by CRs. Most immediately, observations by the upcoming Cherenkov Telescope Array \textsuperscript{29} and Large High Altitude Air Shower Observatory \textsuperscript{30} should both extend the population of $\gamma$-ray-detected SFGs and push existing detections to substantially higher energies. Our model makes clear predictions for both source counts and spectral shapes that can be tested against these data. In the longer term, application of this model to neutrinos will yield predictions that will be testable by IceCube and other neutrino observatories (see the Supplementary Information and Extended Data Figure 3 for further information), and application to synchrotron emission from CR electrons can be used to make predictions for the radio sky that will be testable with the Square Kilometer Array (SKA) and other next-generation radio telescopes. Moreover, because the basis of these predictions is a coherent physical model, rather than just empirical scalings, these predictions can all be made self-consistently.

Supplementary Information is available for this paper.

\newpage

\section*{References}

\begin{enumerate}
\item Ackermann, M. et al. The Spectrum of Isotropic Diffuse Gamma-Ray Emission between 100 MeV and 820 GeV. \textit{Astrophys. J.} \textbf{799}, 86. doi:10.1088/0004-637X/799/1/86 (2015).
\item Cuoco, A., Komatsu, E. \& Siegal-Gaskins, J. M. Joint anisotropy and source count constraints on the contribution of blazars to the diffuse gamma-ray background. \textit{Phys. Rev. D} \textbf{86}, 063004. doi:10.1103/PhysRevD.86.063004 (2012).
\item Di Mauro, M., Calore, F., Donato, F., Ajello, M. \& Latronico, L. Diffuse $\gamma$-Ray Emission from Misaligned Active Galactic Nuclei. \textit{Astrophys. J.} \textbf{780}, 161. doi:10.1088/0004-637X/780/2/161 (2014).
\item Fornasa, M. \& Sánchez-Conde, M. A. The nature of the Diffuse Gamma-Ray Background. \textit{Phys. Rep.} \textbf{598}, 1–58. doi:10.1016/j.physrep.2015.09.002 (2015).
\item Qu, Y., Zeng, H. \& Yan, D. Gamma-ray luminosity function of BL Lac objects and contribution to the extragalactic gamma-ray background. \textit{Mon. Not. R. Astron. Soc.} \textbf{490}, 758–765. doi:10.1093/mnras/stz2651 (2019).
\item Fields, B. D., Pavlidou, V. \& Prodanovi\'c, T. Cosmic Gamma-ray Background from Star-forming Galaxies. \textit{Astro-phys. J. Lett.} \textbf{722}, L199–L203. doi:10.1088/2041-8205/722/2/L199 (2010).
\item Ackermann, M. et al. GeV Observations of Star-forming Galaxies with the Fermi Large Area Telescope. \textit{Astrophys. J.} \textbf{755}, 164. doi:10.1088/0004-637X/755/2/164 (2012).
\item Xia, J.-Q., Cuoco, A., Branchini, E. \& Viel, M. Tomography of the Fermi-LAT $\gamma$-Ray Diffuse Extragalactic Signal via Cross Correlations with Galaxy Catalogs. \textit{Astrophys. J. Suppl.} \textbf{217}, 15. doi:10.1088/0067-0049/217/1/15 (2015).
\item Linden, T. Star-forming galaxies significantly contribute to the isotropic gamma-ray background. \textit{Phys. Rev. D} \textbf{96}, 083001. doi:10.1103/PhysRevD.96.083001 (2017).
\item Ajello, M., Di Mauro, M., Paliya, V. S. \& Garrappa, S. The $\gamma$-Ray Emission of Star-forming Galaxies. \textit{Astrophys. J.} \textbf{894}, 88. doi:10.3847/1538-4357/ab86a6 (2020).
\item Krumholz, M. R. et al. Cosmic ray transport in starburst galaxies. \textit{Mon. Not. R. Astron. Soc.} \textbf{493}, 2817–2833. doi:10.1093/mnras/staa493 (2020).
\item Ajello, M. et al. The Origin of the Extragalactic Gamma-Ray Background and Implications for Dark Matter Annihilation. \textit{Astrophys. J. Lett.} \textbf{800}, L27. doi:10.1088/2041-8205/800/2/L27 (2015).
\item Bell, A. R. Cosmic ray acceleration. \textit{Astropart. Phys.} \textbf{43}, 56–70. doi:10.1016/j.astropartphys.2012.05.022 (2013).
\item Woosley, S. E. \& Weaver, T. A. The evolution and explosion of massive Stars II: Explosive hydrodynamics and nucleosynthesis NASA STI/Recon Technical Report N. 1995.
\item Dermer, C. D. \& Powale, G. Gamma rays from cosmic rays in supernova remnants. \textit{Astron. Astrophys.} \textbf{553}, A34. doi:10.1051/0004-6361/201220394 (2013).
\item Lacki, B. C., Thompson, T. A. \& Quataert, E. The Physics of the Far-infrared-Radio Correlation. I. Calorimetry, Conspiracy,and Implications. \textit{Astrophys. J.} \textbf{717}, 1–28. doi:10.1088/0004-637X/717/1/1 (2010).
\item Bell, A. R. The acceleration of cosmic rays in shock fronts - I. \textit{Mon. Not. R. Astron. Soc.} \textbf{182}, 147–156. doi:10.1093/mnras/182.2.147 (1978).
\item Blandford, R. \& Eichler, D. Particle acceleration at astrophysical shocks: A theory of cosmic ray origin. \textit{Phys. Rep.} \textbf{154}, 1–75. doi:10.1016/0370-1573(87)90134-7 (1987).
\item Caprioli, D. Understanding hadronic gamma-ray emission from supernova remnants. \textit{J. Cosmology Astropart. Phys.} \textbf{2011}, 026. doi:10.1088/1475-7516/2011/05/026 (2011).
\item Caprioli, D. Cosmic-ray acceleration in supernova remnants: non-linear theory revised. \textit{J. Cosmology Astropart. Phys.} \textbf{2012}, 038. doi:10.1088/1475-7516/2012/07/038 (2012).
\item Grogin,  N.  A. et  al. CANDELS: The Cosmic Assembly Near-infrared  Deep Extragalactic Legacy Survey. \textit{Astro-phys. J. Suppl.} \textbf{197}, 35. doi:10.1088/0067-0049/197/2/35 (2011).
\item van der Wel, A. et al. Structural Parameters of Galaxies in CANDELS. \textit{Astrophys. J. Suppl.} \textbf{203}, 24. doi:10.1088/0067-0049/203/2/24 (2012).
\item Ballet, J., Burnett, T. H., Digel, S. W. \& Lott, B. Fermi Large Area Telescope Fourth Source Catalog Data Release 2. arXiv e-prints, arXiv:2005.11208 (2020).
\item Peng, F.-K., Wang, X.-Y., Liu, R.-Y., Tang, Q.-W. \& Wang, J.-F. First Detection of GeV Emission from an Ultraluminous Infrared Galaxy: Arp 220 as seen with the Fermi Large Area Telescope. \textit{Astrophys. J. Lett.} \textbf{821}, L20. doi:10.3847/2041-8205/821/2/L20 (2016).
\item Kornecki, P. et al. The $\gamma$-ray/infrared luminosity correlation of star-forming galaxies. \textit{Astron. Astrophys.} \textbf{641}, A147. doi:10.1051/0004-6361/202038428 (2020).
\item Ando, S., Fornasa, M., Fornengo, N., Regis, M. \& Zechlin, H.-S. Astrophysical interpretation of the anisotropies in the unresolved gamma-ray background. \textit{Phys. Rev. D} \textbf{95}, 123006. doi:10.1103/PhysRevD.95.123006 (2017).
\item Manconi, S. et al. Testing gamma-ray models of blazars in the extragalactic sky. \textit{Phys. Rev. D} \textbf{101}, 103026. doi:10.1103/PhysRevD.101.103026 (2020).
\item Komis, I., Pavlidou, V. \& Zezas, A. Extragalactic gamma-ray background from star-forming galaxies: Will empirical scalings suffice? \textit{Mon. Not. R. Astron. Soc.} \textbf{483}, 4020–4030. doi:10.1093/mnras/sty3354 (2019).
\item Cherenkov Telescope Array Consortium et al. Science with the Cherenkov Telescope Array doi:10.1142/10986 (World Scientific, Singapore, 2019).
\item di Sciascio, G. \& Lhaaso Collaboration. The LHAASO experiment: From Gamma-Ray Astronomy to Cosmic Rays. \textit{Nuc. Part. Phys. Proc.} \textbf{279-281}, 166–173. doi:10.1016/j.nuclphysbps.2016.10.024 (2016).
\end{enumerate}

\newpage

\section*{Methods}
\hspace{\textwidth}\\
Here we describe our methods to compute $\gamma$-ray emission from a single SFG due to both CR ions and leptons, to determine the flux received at Earth from that galaxy, and to apply these models to the CANDELS sample, as well as the details of the Monte Carlo estimation for low redshift source counts.

\subsection*{$\gamma$-ray emission model for CR ions}

In our model, the total rate of $\gamma$-ray emission per unit energy from a SFG is the sum of an ionic component and a leptonic component, $d\dot{N}_\gamma/dE_\gamma = \left.d\dot{N}_\gamma/dE_\gamma\right|_{\rm ion} + \left.d\dot{N}_\gamma/dE_\gamma\right|_{\rm lepton}$. We compute the ionic component as
\begin{equation}
    \left.\frac{d\dot{N}_\gamma}{dE_\gamma}\right|_{\rm ion} = \int_{m_{\rm p}c^2}^{\infty} \left[\frac{1}{\sigma_{\rm pp}} \frac{d\sigma_\gamma}{dE_\gamma}\left(E_{\rm ion}\right)\right] f_{\rm cal}\left(E_{\rm ion}\right) \; \frac{d\dot{N}_{\rm ion}}{dE_{\rm ion}} \; dE_{\rm ion}.
    \label{eq:Lgamma}
\end{equation}
Here $d\dot{N}_{\rm ion}/dE_{\rm ion}$ is the rate per unit energy at which supernovae injection CR ions of energy $E_{\rm ion}$ into the galaxy, $\sigma_{\rm pp} = 40$ mbarn is the mean proton-proton inelastic cross-section, $d\sigma_\gamma/dE_\gamma(E_{\rm ion})$ is the differential cross section for production of $\gamma$-rays of energy $E_\gamma$ by CR ions of energy $E_{\rm ion}$, and $f_{\rm cal}(E_{\rm ion})$ is the calorimetry fraction for CR ions of energy $E_{\rm ion}$. We take $d\sigma_\gamma/dE_\gamma(E_{\rm ion})$ from the parameterised model of Ref.~\textsuperscript{31}. We compute $d\dot{N}_{\rm ion}/dE_{\rm ion}$ from the galactic star formation rate $\dot{M}_*$ by assuming that stars form with a Chabrier initial mass function \textsuperscript{32}, which gives the distribution of masses for newly-formed stars, and that stars with initial mass of $8-50$ M$_\odot$, where M$_\odot$ is the mass of the Sun, end their lives as supernovae \textsuperscript{33}. Each supernova injects $10^{50}$ erg of energy in CR ions \textsuperscript{14,15}, distributed in energy for CR energies $E_{\rm ion} >m_p c^2$ as $d\dot{N}_{\rm ion}/dE_{\rm ion} = \phi \; \dot{M}_* \; \left(p_{\rm ion}/p_0\right)^{-q} dp_{\rm ion}/dE_{\rm ion} \; \exp\left(-E_{\rm ion}/E_{\rm cut}\right)$, where $p_0 = 1$ ${\rm GeV}/c$, the cutoff energy $E_{\rm cut} = 10^8$ GeV, and the spectral index $q=2.2$ \textsuperscript{19,20}. The exact choice of the cutoff energy above $E_{\rm ion} \sim$ PeV makes no practical difference because the injection spectral index $q>2$, so only a small fraction of the total CR energy is injected at $\gtrsim$ PeV energies regardless of $E_{\rm cut}$, and any CRs that are injected at such high energies produce photons that we do not observe due to $\gamma\gamma$ opacity. The normalisation factor $\phi$ that corresponds to our choice of initial mass function, supernova mass range, and CR energy per supernova is $\phi = 7.15 \times 10^{42} \; {\rm s^{-1} \; GeV^{-1} \; M_\odot^{-1} \; yr}$.

The only remaining unknown in \autoref{eq:Lgamma} is the calorimetry fraction $f_{\rm cal}\left(E_{\rm ion}\right)$, which we compute from the recent model of Ref.~\textsuperscript{11}. The basic premise of the model is that, in the neutral phase that dominates the mass of the ISM and thus the set of available targets for $\gamma$-ray production, CR transport is primarily by streaming along magnetic field lines. However, this yields approximately diffusive transport when averaged over scales comparable to or larger than the coherence length of the magnetic field, with a diffusion coefficient $D \approx V_{\rm st} h_g/M_A^3$, where $V_{\rm st}$ is the CR streaming speed, $h_g$ is the gas scale height, and $M_A$ is the Alfv\'en Mach number of the turbulence. For diffusive transport with losses in a disc geometry, the calorimetry fraction is given by (using the favoured parameters of Ref.~\textsuperscript{11})
\begin{equation}
    f_{\rm cal}(E_{\rm ion}) = 1 - \left[ {}_0F_1 \left( \frac{1}{5}, \frac{16}{25}\tau_{\text{eff}} \right) + \frac{3 \; \tau_{\text{eff}}}{4 \; M_A^3} \; {}_0F_1 \left( \frac{9}{5}, \frac{16}{25}\tau_{\text{eff}} \right) \right]^{-1},
\end{equation}
where $_{0}F_1$ is the generalised hypergeometric function and $\tau_{\rm eff}$ is the dimensionless effective optical depth of the ISM, given by
\begin{equation}
    \tau_{\rm eff} = \frac{\sigma_{\rm pp} \; \eta_{\rm pp} \; \Sigma_g \; h_g \; c}{2 \; D_0 \; \mu_{\rm p} \; m_{\rm H}}.
\end{equation}
Here $\eta_{\rm pp}=0.5$ is the elasticity of $pp$ collisions, $\Sigma_g$ is the gas surface density of the galactic disc, $c$ is the speed of light, $D_0$ is the diffusion coefficient at the galactic midplane, $\mu_{\rm p} = 1.17$ is the number density of nucleons per proton, and $m_{\rm H} = 1.67 \times 10^{-24}$ g is the mass of a hydrogen atom.

To evaluate the calorimetry fraction for a CR of energy $E_{\rm ion}$, we must therefore determine the midplane diffusion coefficient $D_0$ for CRs of that energy, which in turn depends on the streaming speed $V_{\rm st}$. This speed is dictated by the balance between excitation of the streaming instability and dissipation of the instability by ion-neutral damping, the dominant dissipation mechanism in the weakly-ionised neutral ISM. Balancing these two effects yields a CR proton streaming velocity
\begin{equation}
V_{\rm st} \approx \min\left[c,
V_{\rm Ai} \left(
1 + \frac{\gamma_d \; \chi \; M_A \; c \; \rho^{3/2}}{4 \; \pi^{1/2} \; C \; e \; u_{\rm LA} \; \mu_i \; \gamma^{-q+1}}
\right)
\right]
\label{eq:streaming_speed}
\end{equation}
where $V_{\rm Ai}$ is the ion Alfv\'en speed, $\gamma_d = 4.9\times 10^{13}$ cm$^3$ g$^{-1}$ is the ion-neutral drag coefficient, $\chi$ is the ionised mass fraction, $\rho = \Sigma_g/2 h_g$ is the midplane mass density of the ISM, $C$ is the midplane number density of CRs, $e$ is the elementary charge, $u_{\rm LA}$ is the velocity dispersion of Alfv\'en modes in the ISM at the outer scale of the turbulence, $\mu_i$ is the atomic mass of the dominant ion species, $q$ is the index of the CR energy distribution, and $\gamma = E_{\rm ion}/m_p c^2$ is the Lorentz factor of the CR. Since $\gamma$-ray production in our model is dominated by galaxies with high star formation rates and gas surface densities within which i) the ISM is molecule-dominated, ii) the main ionised species is C$^+$, and iii) the magnetic field is set by a turbulent dynamo, we adopt the fiducial parameters of Ref.~\textsuperscript{11} appropriate for such galaxies. Specifically, we take $\chi = 10^{-4}$, $\mu_i = 12$, $M_A = 2$, $V_{\rm Ai} = u_{\rm LA}/\chi^{1/2} M_A$, and $u_{\rm LA} = \sigma_g/\sqrt{2}$, where $\sigma_g$ is the gas velocity dispersion of the galaxy. We also adopt $q=2.2$, consistent with our assumed injection spectrum. We have chosen this set of parameters without any fine tuning, by selecting values that are generally accepted as being the most appropriate for the type of source that dominates the emission. However, we explore the parameter space in the Supplementary Information and show a selection of results in Extended Data Figure 1.

At this point we have specified all the ingredients required to compute $f_{\rm cal}(E_{\rm ion})$ for a galaxy of known $\Sigma_g$, $\sigma_g$, and $h_g$, save one: $C$, the CR number density. We estimate this as follows: consistent with our discussion above, for a galaxy with star formation rate per unit area $\dot{\Sigma}_*$, the CR ion injection rate per unit area is
\begin{equation}
\begin{split}
    \frac{d\dot{N}_{\rm ion}}{dA} &= \phi \dot{\Sigma}_* \int_{m_p c^2}^\infty \left(\frac{p_{\rm ion}}{p_0}\right)^{-q} \frac{E_{\rm ion}}{p_{\rm ion}} \; e^{-E_{\rm ion}/E_{\rm cut}} \, dE_{\rm ion}\\ &\approx 2.61 \times 10^{43} \; \dot{\Sigma}_* \; {\rm s^{-1} M_\odot^{-1} yr}.
\end{split}
\end{equation}
The CR number density at the midplane is then given by
\begin{equation}
    C \approx \frac{t_{\rm loss}}{2 h_g} \left(\frac{d \dot{N}_{\rm ion}}{dA}\right),
\end{equation}
where $t_{\rm loss}$ is the CR loss time. This is given by $t_{\rm loss} = 1/\left( t_{\rm col}^{-1} + t_{\rm diff}^{-1} \right)$, where the timescale for losses in inelastic hadronic collisions is $t_{\text{\rm col}} = 1/\left( \rho \sigma_{\rm pp} \eta_{\rm pp} c/\mu_{p} m_{\rm H} \right)$ and the diffusive escape time $t_{\text{\rm diff}} =h_g^2 D_0^{-1}$. For the systems with high gas densities and high star formation rate that dominate $\gamma$-ray production, the loss time is generally dominated by collisional losses. Conversely,  for systems forming stars more sedately and with lower density environments, it is generally determined by the diffusive escape time.

As a final note, we point out that the model of Ref.~\textsuperscript{11} applies at CR energies up to tens of TeV in galaxies whose interstellar media are mainly molecular (most galaxies with star formation rates above a few Solar masses per year), but may break down above tens of GeV in galaxies where the gas is mostly atomic \textsuperscript{34}. Thus the model might not predict the correct degree of calorimetry at $\gtrsim 100$ GeV energies in dominantly atomic galaxies. As shown in Extended Data Figure 2, however, low star formation rate galaxies make only a small contribution to the background, and thus a possible error in them will have minimal effects on the final result.

We illustrate the behaviour of $f_{\rm cal}\left(E_{\rm ion}\right)$ over a range of gas surface densities and redshifts as applied to the CANDELS sample (see below) in Extended Data Figures 4 and 5. 

\subsection*{$\gamma$-ray emission model for CR leptons}

CR electrons and positrons (since both behave identically, for brevity we just write electrons, but everything that follows should be understood as applying to a mix of both) are either injected directly by diffusive shock acceleration in supernova remnants at the same time as CR ions (primary production), or appear in the decay of charged pions $\pi^\pm$ produced by collisions of CR ions with the ISM (secondary production). We assume the former carry a total energy equal to 2\% of supernova kinetic energy \textsuperscript{16}, and have the same injection spectrum as the CR ions, $d\dot{N}_{\rm e}/dE_{\rm e}|_{1} \propto p_{\rm e}^{-q} (E_{\rm e}/p_{\rm e}) e^{-E_{\rm e}/E_{\rm cut,e}}$, with $q = 2.2$, but a lower spectral cut-off energy at $E_{\rm cut,e} = 100$ TeV \textsuperscript{35}. For the latter we follow Refs.~\textsuperscript{36,37}: we first compute the rate at which CR ions produce pions of energy $E_\pi$,
\begin{equation}
    \frac{d\dot{N}_{\pi}}{dE_\pi} = \frac{n_{\rm H} \; c}{K_{\rm \pi}} \; \beta \; \sigma_{\rm pp}(E_{\rm ion}) \; \frac{d\dot{N}_{\rm ion}}{dE_{\rm ion}} f_{\rm cal}\left(E_{\rm ion}\right),
\end{equation}
where $n_{\rm H} = \Sigma_g/(2\mu_p m_{\rm H} h_g)$ is the ISM number density, $c$ the speed of light, $\beta$ is the CR velocity divided by $c$, $E_\pi = K_\pi (E - m_p c^2)$, and $K_{\rm \pi} = 0.17$ is the fraction of energy transferred from the CR to the pion. Then the rate at which these pions produce secondary electrons is
\begin{equation}
\frac{d\dot{N}_{\rm e}}{dE_{\rm e}}\bigg|_2 =  2 \int_{E_{\rm e}/E_\pi}^{1}   f_{\rm \nu_{\mu}^{\left(2\right)}}\left( x \right) \; \frac{d\dot{N}_\pi}{dE_\pi}\left( \frac{E_{\rm {\rm e}}}{x} \right) \; \frac{dx}{x},
\label{eq:2ndlep_spec}
\end{equation}
where $f_{\rm \nu_{\mu}^{\left(2\right)}}\left( x \right)$ is a dimensionless fititng function given in Ref.~\textsuperscript{36}. Thus the total CR electron injection rate is $d\dot{N}_e/dE_e = d\dot{N}_e/dE_e|_1 + d\dot{N}_e/dE_e|_2$.

The electrons are subject to four dominant loss processes: collisional ionisation, synchrotron radiation, bremsstrahlung, and inverse Compton scattering; as discussed in the main text, diffusive escape from the galaxy is negligible in comparison. We adopt the following parameterisations from the literature for the total energy loss rates $dE_e/dt$ for electrons of energy $E_e$:
\begin{eqnarray}
-\frac{dE_{\rm e}}{dt}\bigg|_{\rm ion} & = & \frac{ 9 }{ 4 } \sigma_T c \; m_{\rm e} c^2 \; n_{\rm H} \left[\ln\gamma + \frac{2}{3} \ln\left(\frac{m_{\rm e} c^2}{15 {\rm eV}}\right)\right] \\
-\frac{dE_{\rm e}}{dt}\bigg|_{\rm sync} & = & \frac{1}{6 \pi} \sigma_T c \; B^2 \; \gamma^2 \beta^2 \\
-\frac{dE_{\rm e}}{dt}\bigg|_{\rm brems} & = & \frac{3}{\pi}\alpha \sigma_T c m_c c^2 \gamma n_{\rm H}
\left\{
\begin{array}{ll}
     \ln \gamma + \ln 2 - 1/3 & \gamma \lesssim 15  \\
     \Phi_{1,\rm H}(1/4\alpha\gamma)/8 & \gamma \gtrsim 15 
\end{array}
\right. \\
-\frac{dE_{\rm e}}{dt}\bigg|_{\rm IC} & = & \frac{20}{\pi^4}\sigma_T c \gamma^2 u_{\rm rad} Y(4\gamma E_{\rm peak}/m_e c^2).
\end{eqnarray}
In these expressions, $\sigma_T$ is the Thomson cross section, $\alpha$ is the fine structure constant, $m_{\rm e}$ is the electron mass, $\gamma = E_c/m_{\rm e} c^2$ is the electron Lorentz factor, $E_{\rm peak}$ is the energy where the infrared background peaks (derived from the dust temperature $T_{\rm dust} = 98 \left(1+z\right)^{-0.065}+6.9\log{\dot{M}_*/M_*}$ \textsuperscript{38} and injected as a diluted modified black body spectrum \textsuperscript{39} which peaks in photon number at $E_{\rm peak} = 2.82 \; k_{\rm B} T_{\rm dust}$ where $k_{\rm B}$ is the Boltzmann constant), $B = V_{\rm Ai} /\sqrt{4\pi n_{\rm H} \mu_p m_{\rm H} \chi}$ is the magnetic field strength, $u_{\rm rad}$ is the radiation energy density (which based on empirical measurements in nearby galaxies we set equal to the magnetic energy density \textsuperscript{40}, $u_{\rm rad} = u_{\rm mag} = B^2/8\pi$), and $\Phi_{1,\rm H}$ and $Y$ are dimensionless numerical fitting functions; these expressions are taken from Refs.~\textsuperscript{41} (ionisation), \textsuperscript{42} (synchrotron), \textsuperscript{41} (bremsstrahlung), and \textsuperscript{43} (inverse Compton), and the definitions of the fitting functions are given in those references. Given these loss rates, the steady-state spectrum in the galaxy is given approximately by
\begin{equation}
\frac{dN_{\rm e}}{dE_{\rm e}} \simeq \frac{d\dot{N}_{\rm e}}{dE_{\rm e}} \; t_{\rm loss}\left(E_{\rm e}\right),
\end{equation}
where $t_{{\rm loss},i} = E_{\rm e}/\frac{dE_{\rm e}}{dt} \big|_i$ is the loss time due to process $i$, and $t_{\rm loss} = \left( \sum_i t_{{\rm loss},i}^{-1} \right)^{-1}$ is the total loss time.

Of the four loss processes, only bremsstrahlung and inverse Compton scattering produce $\gamma$-rays. We compute the resulting emission using expressions analogous to \autoref{eq:Lgamma}. For bremsstrahlung,
\begin{equation}
\frac{d\dot{N}_{\gamma}}{dE_{\gamma}}\bigg|_{\rm brems} = c \; \frac{n_{\rm H}}{E_\gamma} \; \int_{E_\gamma}^\infty \sigma_{\rm BS}\left( E_\gamma, E_{\rm e} \right) \; \frac{dN_{\rm e}}{dE_{\rm e}} \; dE_{\rm e},
\label{eq:brems}
\end{equation}
where $\sigma_{\rm BS}$ is the cross section for production of photons of energy $E_\gamma$ by electrons of energy $E_{\rm e}$; we take our expression for $\sigma_{\rm BS}$ from Refs.~\textsuperscript{41,44,45}. Similarly, for inverse Compton we have \textsuperscript{35,46}
\begin{equation}
    \frac{d\dot{N}_{\gamma}}{dE_{\gamma}}\bigg|_{\rm IC} = \frac{3}{4} \sigma_{\rm T} c \; \frac{u_{\rm rad}}{E_{\rm peak}^2} \; \int_{E_{\rm e, min}}^\infty \frac{dN_{\rm e}}{dE_{\rm e}} \; \frac{G\left(a, \Gamma\right)}{\gamma^2} \; dE_{\rm e},
    \label{eq:IC}
\end{equation}
where $G(a,\Gamma) = 2 a \ln a + (1+2a)(1-a) + (\Gamma a)^2(1-a)/2(1+\Gamma a)$, $\Gamma = 4E_{\rm peak} E_e/m_e^2 c^4$, and $a=E_\gamma/\Gamma (E_e-E_\gamma)$. The total leptonic contribution to $\gamma$-ray production is simply $\left.d\dot{N}_\gamma/dE_\gamma\right|_{\rm lepton} = \left.d\dot{N}_\gamma/dE_\gamma\right|_{\rm brmes} + \left.d\dot{N}_\gamma/dE_\gamma\right|_{\rm IC}$.

We show the contribution of leptonic emission to the diffuse, isotropic $\gamma$-ray background divided by emission mechanism, and by primary versus secondary, in Extended Data Figure 6.

\subsection*{$\gamma$-ray flux at Earth}

To obtain the total observed $\gamma$-ray flux for a galaxy, we must account for the attenuation of $\gamma$-ray photons by galactic far-infrared and extragalactic background light (EBL) photons. We compute the optical depth $\tau_{\gamma\gamma}$ due to the former using the model of Ref.~\textsuperscript{47}, and the optical depth from the latter, $\tau_{\rm EBL}$, using the model of Ref.~\textsuperscript{48}. Taking these into account, we can compute the specific photon flux $dF_\gamma/dE_\gamma$ (i.e., the number of photons per unit area, time and energy) received at the Earth from a galaxy at redshift $z$ as
\begin{equation}
    \frac{dF_\gamma}{dE_\gamma} = \frac{\left(1+z\right)^2}{4 \pi \; d_{\rm L}^2\left(z\right)} \; \left.\frac{d\dot{N}_\gamma}{dE_\gamma}\right|_{E_\gamma \left(1+z\right)} \; e^{-\tau_{\rm EBL}\left( E_{\rm \gamma}, \; z \right)} \; \; e^{-\tau_{\rm \gamma\gamma}\left( E_{\rm \gamma} \left(1+z\right)\right)}
    \label{eq:specgamma+opacity}
\end{equation}
where $\left.d\dot{N}_\gamma/dE_\gamma\right|_{E_\gamma(1+z)}$ is the total $\gamma$-ray production from both CR ions and electrons evaluated at an energy $E_\gamma(1+z)$, and $d_{\rm L}\left(z\right)$ is the luminosity distance of the source. 

The radiation that is absorbed by the host galaxy and extragalactic photon fields is reprocessed to lower energies in the pair-production cascade. We parameterise the photon spectrum from this cascade using the method developed in Ref.~\textsuperscript{49}. 
For the purposes of our calculation here, we include the effect of the cascade by adding a component to $dF_\gamma/dE_\gamma$ with a spectral shape as computed by Ref.~\textsuperscript{49}, and with a normalisation such that its energy is equal to the integrated energy lost to photon-photon scattering.

\subsection*{Application of the model to CANDELS galaxies}

We apply our model to each individual galaxy in the CANDELS sample. The full sample contains 34,930 galaxies, but we exclude those whose parameters are uncertain because they contain bright active galactic nuclei, have unreliable redshifts, or lack a good fit to the surface brightness profile. This leaves a sample of 22,279 galaxies.

Our calorimetry model requires, as input, the gas surface density $\Sigma_g$, scale height $h_g$, and velocity dispersion $\sigma_g$, along with the surface density of star formation $\dot{\Sigma}_*$. However, the CANDELS data set in Ref.~ \textsuperscript{22}, that we use, provides only the cosmological redshift $z$, stellar mass $M_*$, half-light or effective radius $R_{\rm e}$ (corrected to 5000 \AA~according to Ref.~\textsuperscript{50}), and total star formation rate $\dot{M}_*$ for our sample galaxies. We must therefore estimate the gas properties from observed correlations between gas and stellar properties. We do so as follows.

The half light radius $R_{\rm e}$ at $5000$ \AA~serves as a first order estimate of how the star formation and matter are distributed throughout the galactic disc. We therefore  estimate the star formation rate surface density as $\dot{\Sigma}_* = \dot{M}_*/2 \pi R_{\rm e}^2$ and the stellar surface density as $\Sigma_* = M_*/2 \pi R_{\rm e}^2$. We estimate the gas surface density from the observed correlation between gas, stellar, and star formation surface densities given by Ref.~\textsuperscript{51}:
\begin{equation}
\frac{\Sigma_g}{\rm M_\odot \; pc^{-2}} = 10^{10.28} \; \frac{\dot{\Sigma}_{*}}{\rm M_\odot \; yr^{-1} \; pc^{-2}} \; \left(\frac{\Sigma_{*}}{{\rm M_\odot \; pc^{-2}}}\right)^{-0.48}
\end{equation}
Similarly, there is a strong correlation between galaxy star formation rates and velocity dispersions, which we use to derive $\sigma_g$. For this purpose we fit the relationship using the MaNGA galaxy sample \textsuperscript{52}. A powerlaw fit to the data obtained in this survey (Fig.~6 of Ref.~\textsuperscript{52}) gives
\begin{equation}
\frac{\sigma_g}{\rm km \; s^{-1}} = 32.063 \left( \frac{\dot{M}_*}{\rm M_\odot \; yr^{-1}} \right)^{0.096}
\end{equation}
Finally, we derive the gas scale height under the assumption that the gas is in vertical hydrostatic equilibrium, in which case the scale height is \textsuperscript{53}
\begin{equation}
h_g = \frac{\sigma_g^2}{\pi G \left(\Sigma_{g} + \Sigma_{*}\right)}
\end{equation}

With these gas properties in hand, we calculate an observed spectrum for each galaxy in the CANDELS sample by using \autoref{eq:specgamma+opacity}, and we then sum over the sample to predict the $\gamma$-ray flux per unit energy per unit solid angle $\Phi(E_\gamma)$ produced by SFGs. In practice, we compute the sum as:
\begin{equation}
    \Phi\left( E_\gamma \right) = \frac{1}{\Omega_{\rm S}} \sum_{j=1}^{n_{\rm zbin}} f_{\rm corr, j} \sum_{i=1}^{n_{\rm S, j}} \left(\frac{dF_{\gamma, i}}{dE_\gamma} \right)_{i,j}
\end{equation}
Here $\Omega_{\rm S} = 173 \; {\rm arcmin}^2$ is the solid angle surveyed by CANDELS, $n_{\rm S, j}$ is the number of surveyed galaxies in the $j$th redshift bin, $(dF_\gamma/dE_\gamma)_{i,j}$ is the flux from the $i$th galaxy in this bin predicted using \autoref{eq:specgamma+opacity}, and $n_{\rm zbin}$ the number of redshift bins. The factor $f_{\rm corr, j}$ is the ratio of the expected total star formation rate in each redshift bin (based on the measured cosmic star formation history \textsuperscript{54} and obtained by integrating the star formation rate density over the volume in each redshift bin) to the sum of the star formation rates of CANDELS galaxies in that bin; its purpose is to correct for the fact that, due to its limited field of view and various observational biases, the distribution of star formation with respect to redshift in CANDELS does not precisely match the total star formation history of the Universe.
We use redshift bins of size $\Delta z = 0.1$, chosen to ensure that the number of sample galaxies in each bin is large enough that the uncertainty in the mean spectrum due to Poisson sampling of the galaxy population is small.

For the purposes of constructing Figure 2, we also require not just the flux, but the total $\gamma$-ray luminosity in the \textit{Fermi} band. We compute this by integrating $E_\gamma \, d\dot{N}_\gamma/dE_\gamma$ (computed as the sum of Equations \ref{eq:Lgamma}, \ref{eq:brems}, and \ref{eq:IC}) from $E_\gamma = 0.1-100$ GeV. Since this comparison also requires the far-infrared luminosity, we convert the star formation rate to an far-infrared luminosity in the $8-1000$ $\mu$m band using the relation in Refs.~\textsuperscript{7,55}, corrected to a Chabrier IMF \textsuperscript{56}; this conversion is valid for star formation rates $\gtrsim 1$ $M_\odot$ yr$^{-1}$, which encompasses almost all of the observed sample to which we wish to compare.

\subsection*{Monte Carlo estimation for the local population}

To estimate the observable source count distribution for \textit{Fermi} LAT, as shown in Figure 3, we cannot use the CANDELS catalogue directly, because CANDELS has a narrow field of view that provides very little sampling of galaxies at $z \lesssim 0.1$, whereas these local sources are the only ones \textit{Fermi} LAT can resolve. We therefore use a Monte Carlo scheme to simulate a nearby, low-redshift ($z<0.1$) galaxy population that follows the observed distribution of star formation rates in the local Universe to account for cosmic variance, and where the correlation between galaxy star formation rate and $\gamma$-ray luminosity is the same as what our model predicts for the low redshift ($z<1.5$) part of the CANDELS sample. 

The first step is to produce a sample of SFGs. To do so, we draw galaxies from the observed distribution of star formation rates in the local Universe \textsuperscript{57,58}. For each SFG drawn, we also draw an associated redshift in the range $z=0 - 0.1$, with probability proportional to the co-moving volume element. We continue drawing galaxies until the total star formation rate of the population we have drawn matches the integrated star formation rate within the volume $z=0 - 0.1$ as determined from the cosmic star formation history \textsuperscript{54}. The second step is to assign $\gamma$-ray luminosities for these galaxies based on our model for the CANDELS galaxies. For this purpose, we apply our model to predict the photon luminosity integrated over the 1 - 100 GeV band (i.e., the number of photons per unit time emitted in this energy range) for all CANDELS galaxies with $z<1.5$, and fit a power law relationship between this luminosity and the star formation rate; we neglect $\gamma\gamma$ opacity in this calculation, since this effect is unimportant for the galaxies at $z < 0.1$ and the energy range $<100$ GeV that we are simulating. We then assign each of our SFGs a $\gamma$-ray photon luminosity using this powerlaw fit, and in conjunction with the redshift, an observed photon flux $S$. 

At this point we have a sample of $\gamma$-ray photon fluxes $S$ for simulated $z<0.1$ SFGs, which we can place in bins of $S$ to construct a synthetic prediction for $S^2(dN/dS)$. We carry out 13,000 Monte Carlo trials of this type, and in each bin of $S$ record the mean and the 68\% and 90\% probability intervals, which we show as the blue points and bands in Figure 3. Our method for computing the analogous confidence intervals for the observations is described in the Supplementary Information.

\newpage

\section*{Methods References}

\begin{enumerate}
\setcounter{enumi}{30}
\item Kafexhiu, E., Aharonian, F., Taylor, A. M. \& Vila, G. S. Parametrization of gamma-ray production cross sections for pp interactions in a broad proton energy range from the kinematic threshold to PeV energies. \textit{Phys. Rev. D} \textbf{90}, 123014. doi:10.1103/PhysRevD.90.123014 (2014).
\item Chabrier, G. Galactic Stellar and Substellar Initial Mass Function. \textit{Publ. Astron. Soc. Pac.} \textbf{115}, 763–795. doi:10.1086/376392 (2003).
\item Heger, A., Fryer, C. L., Woosley, S. E., Langer, N. \& Hartmann, D. H. How Massive Single Stars End Their Life. \textit{Astrophys. J.} \textbf{591}, 288–300. doi:10.1086/375341 (2003).
\item Crocker, R. M., Krumholz, M. R. \& Thompson, T. A. Cosmic rays across the star-forming galaxy sequence - I. Cosmic ray pressures and calorimetry. \textit{Mon. Not. R. Astron. Soc.} \textbf{502}, 1312–1333. doi:10.1093/mnras/stab148 (2021).
\item Peretti, E., Blasi, P., Aharonian, F. \& Morlino, G. Cosmic ray transport and radiative processes in nuclei of starburst galaxies. \textit{Mon. Not. R. Astron. Soc.} \textbf{487}, 168–180. doi:10.1093/mnras/stz1161 (2019).
\item Kelner, S. R., Aharonian, F. A. \& Bugayov, V. V. Energy spectra of gamma rays, electrons, and neutrinos produced at proton-proton interactions in the very high energy regime. \textit{Phys. Rev. D} \textbf{74}, 034018. doi:10.1103/PhysRevD.74.034018 (2006).
\item Kelner, S. R., Aharonian, F. A. \& Bugayov, V. V. Erratum: Energy spectra of gamma rays, electrons, and neutrinos produced at proton-proton interactions in the very high energy regime [Phys. Rev. D 74, 034018 (2006)]. \textit{Phys. Rev. D} \textbf{79}, 039901. doi:10.1103/PhysRevD.79.039901 (2009).
\item Magnelli, B. et al. The evolution of the dust temperatures of galaxies in the $\rm SFR-M_*$ plane up to z ~ 2. \textit{Astron. Astrophys.} \textbf{561}, A86. doi:10.1051/0004-6361/201322217 (2014).
\item Persic, M., Rephaeli, Y. \& Arieli, Y. Very-high-energy emission from M 82. \textit{Astron. Astrophys.} \textbf{486}, 143–149. doi:10.1051/0004-6361:200809525 (2008).
\item Thompson, T. A. Gravitational Instability in Radiation Pressure-Dominated Backgrounds. \textit{Astrophys. J.} \textbf{684}, 212–225. doi:10.1086/589227 (2008).
\item Schlickeiser, R. Cosmic Ray Astrophysics (Springer, Heidelberg, 2002).
\item Ghisellini, G. Radiative Processes in High Energy Astrophysics doi:10.1007/978-3-319-00612-3 (Springer,Heidelberg, 2013).
\item Fang, K., Bi, X.-J., Lin, S.-J. \& Yuan, Q. Klein-Nishina effect and the cosmic ray electron spectrum. arXiv e-prints,arXiv:2007.15601 (2020).
\item Bethe, H. \& Heitler, W. On the Stopping of Fast Particles and on the Creation of Positive Electrons. \textit{Proc. R. Soc. Lon. S. A} \textbf{146}, 83–112. doi:10.1098/rspa.1934.0140 (1934).
\item Gould, R. J. High-Energy Bremsstrahlung in Collisions of Electrons with One- and Two-Electron Atoms. \textit{Phys. Rev.} \textbf{185}, 72–79. doi:10.1103/PhysRev.185.72 (1969).
\item Blumenthal, G. R. \& Gould, R. J. Bremsstrahlung, Synchrotron Radiation, and Compton Scattering of High-Energy Electrons Traversing Dilute Gases. \textit{Rev. Mod. Phys.} \textbf{42}, 237–271. doi:10.1103/RevModPhys.42.237 (1970).
\item Razzaque, S., Mészáros, P. \& Zhang, B. GeV and Higher Energy Photon Interactions in Gamma-Ray Burst Fireballs and Surroundings. \textit{Astrophys. J.} \textbf{613}, 1072–1078. doi:10.1086/423166 (2004).
\item Franceschini, A. \& Rodighiero, G. The extragalactic background light revisited and the cosmic photon-photon opacity. \textit{Astron. Astrophys.} \textbf{603}, A34. doi:10.1051/0004-6361/201629684 (2017).
\item Berezinsky, V. \& Kalashev, O. High-energy electromagnetic cascades in extragalactic space: Physics and features. \textit{Phys. Rev. D} \textbf{94}, 023007. doi:10.1103/PhysRevD.94.023007 (2016).
\item van der Wel, A. et al. 3D-HST+CANDELS: The Evolution of the Galaxy Size-Mass Distribution since z = 3. \textit{Astrophys. J.} \textbf{788}, 28. doi:10.1088/0004-637X/788/1/28 (2014).
\item Shi, Y. et al. Extended Schmidt Law: Role of Existing Stars in Current Star Formation. \textit{Astrophys. J.} \textbf{733}, 87. doi:10.1088/0004-637X/733/2/87 (2011).
\item Yu, X. et al. What drives the velocity dispersion of ionized gas in star-forming galaxies? \textit{Mon. Not. R. Astron. Soc.} \textbf{486}, 4463–4472. doi:10.1093/mnras/stz1146 (2019).
\item Forbes, J., Krumholz, M. \& Burkert, A. Evolving Gravitationally Unstable Disks over Cosmic Time: Implications for Thick Disk Formation. \textit{Astrophys. J.} \textbf{754}, 48. doi:10.1088/0004-637X/754/1/48 (2012).
\item Madau, P. \& Dickinson, M. Cosmic Star-Formation History. \textit{Ann. Rev. Astron. Astrophys.} \textbf{52}, 415–486. doi:10.1146/annurev-astro-081811-125615 (2014).
\item Kennicutt Robert C., J. The Global Schmidt Law in Star-forming Galaxies. \textit{Astrophys. J.} \textbf{498}, 541–552. doi:10.1086/305588 (1998).
\item Crain, R. A., McCarthy, I. G., Frenk, C. S., Theuns, T. \& Schaye, J. X-ray coronae in simulations of disc galaxy formation. \textit{Mon. Not. R. Astron. Soc.} \textbf{407}, 1403–1422. doi:10.1111/j.1365-2966.2010.16985.x (2010).
\item Bothwell, M. S. et al. The star formation rate distribution function of the local Universe. \textit{Mon. Not. R. Astron. Soc.} \textbf{415}, 1815–1826. doi:10.1111/j.1365-2966.2011.18829.x (2011).
\item Bothwell, M. S. et al. Erratum: The star formation rate distribution function of the local Universe. \textit{Mon. Not. R. Astron. Soc.} \textbf{438}, 3608–3608. doi:10.1093/mnras/stt2458 (2014).
\item Rodríguez Zaurín, J., Tadhunter, C. N. \& González Delgado, R. M. Optical spectroscopy of Arp220: the star formation history of the closest ULIRG. \textit{Mon. Not. R. Astron. Soc.} \textbf{384}, 875–885. doi:10.1111/j.1365-2966.2007.12658.x (2008).
\item Wright, G. S., James, P. A., Joseph, R. D., McLean, I. S. \& Doyon, R. Infrared images of merging galaxies. in NASA Conference Publication (eds Sulentic, J. W., Keel, W. C. \& Telesco, C. M.) \textbf{3098} (1990), 321–326.
\item Parkash, V., Brown, M. J. I., Jarrett, T. H. \& Bonne, N. J. Relationships between HI Gas Mass, Stellar Mass, and the Star Formation Rate of HICAT+WISE (H I-WISE) Galaxies. \textit{Astrophys. J.} \textbf{864}, 40. doi:10.3847/1538-4357/aad3b9
(2018).
\item IRSA. NASA/IPAC Infrared Science Archive https://irsa.ipac.caltech.edu.
\item Sakamoto, K. et al. Molecular Superbubbles in the Starburst Galaxy NGC 253. \textit{Astrophys. J.} \textbf{636}, 685–697. doi:10.1086/498075 (2006).
\item Bendo, G. J. et al. ALMA observations of 99 GHz free-free and H40$\alpha$ line emission from star formation in the centre of NGC 253. \textit{Mon. Not. R. Astron. Soc.} \textbf{450}, L80–L84. doi:10.1093/mnrasl/slv053 (2015).
\item Rahmani, S., Lianou, S. \& Barmby, P. Star formation laws in the Andromeda galaxy: gas, stars, metals and the surface density of star formation. \textit{Mon. Not. R. Astron. Soc.} \textbf{456}, 4128–4144. doi:10.1093/mnras/stv2951 (2016).
\item Iijima, T., Ito, K., Matsumoto, T. \& Uyama, K. Near-Infrared Profile of M31. \textit{Pub. Astron. Soc. Japan} \textbf{28}, 27–34 (1976).
\item Braun, R. \& Walterbos, R. A. M. Physical Properties of Neutral Gas in M31 and the Galaxy. \textit{Astrophys. J.} \textbf{386}, 120. doi:10.1086/170998 (1992).
\item Caldú-Primo, A. \& Schruba, A. Molecular Gas Velocity Dispersions in the Andromeda Galaxy. \textit{Astron. J.} \textbf{151}, 34. doi:10.3847/0004-6256/151/2/34 (2016).
\item Davis, B. L., Graham, A. W. \& Cameron, E. Black Hole Mass Scaling Relations for Spiral Galaxies. II. $M_{\rm BH-M_{\star,tot}}$ and $M_{\rm BH-M_{\star,disk}}$. \textit{Astrophys. J.} \textbf{869}, 113. doi:10.3847/1538-4357/aae820 (2018).
\item de Vaucouleurs, G. Southern Galaxies. V. Isophotometry of the Large Barred Spiral NGC 4945. \textit{Astrophys. J.} \textbf{139}, 899. doi:10.1086/147824 (1964).
\item Bendo, G. J. et al. Free-free and H42$\alpha$ emission from the dusty starburst within NGC 4945 as observed by ALMA. \textit{Mon. Not. R. Astron. Soc.} \textbf{463}, 252–269. doi:10.1093/mnras/stw1659 (2016).
\item Ott, M., Whiteoak, J. B., Henkel, C. \& Wielebinski, R. Atomic and molecular gas in the starburst galaxy NGC 4945. \textit{Astron. Astrophys.} \textbf{372}, 463–476. doi:10.1051/0004-6361:20010505 (2001).
\item Stettner, J. Measurement of the diffuse astrophysical muon-neutrino spectrum with ten years of IceCube data. in 36th International Cosmic Ray Conference (ICRC2019) \textbf{36} (2019), 1017.
\end{enumerate}

\newpage

\subsection*{Acknowledgements}

This research has made use of the NASA/IPAC Infrared Science Archive, which is funded by the National Aeronautics and Space Administration and operated by the California Institute of Technology. Funding for this work was provided by the Australian Government through the Australian Research Council, awards FT180100375 (MRK) and DP190101258 (RMC and MRK), and the Australian National University through a research scholarship (MAR). RMC thanks Oscar Macias and Shin'ichiro Ando for enlightening conversations while a Kavli IPMU-funded guest of the GRAPPA Institute at the University of Amsterdam.

\subsection*{Author contributions}
All authors were involved in the design of the study and the interpretation of the results. MAR performed the modelling and data analysis with input from MRK, RMC, and SC. The manuscript was written by MAR, MRK and RMC, and reviewed by all authors.

\subsection*{Data availability}
The data that were used to produce the Figures and that support the findings of this study are available in Zenodo with the identifier 10.5281/zenodo.4764111

\subsection*{Code availability}
The code used to derive the key findings of this study is available in Zenodo with the identifier 10.5281/zenodo.4609628

\subsection*{Author information}
The authors declare no competing interests, financial or otherwise. Correspondence and requests for materials should be addressed to the first author. Reprints and permissions information is available at www.nature.com/reprints.

\newpage

\renewcommand{\figurename}{Extended Data Table}
\setcounter{figure}{0}

\begin{figure*}
  \includegraphics[width=0.99\textwidth]{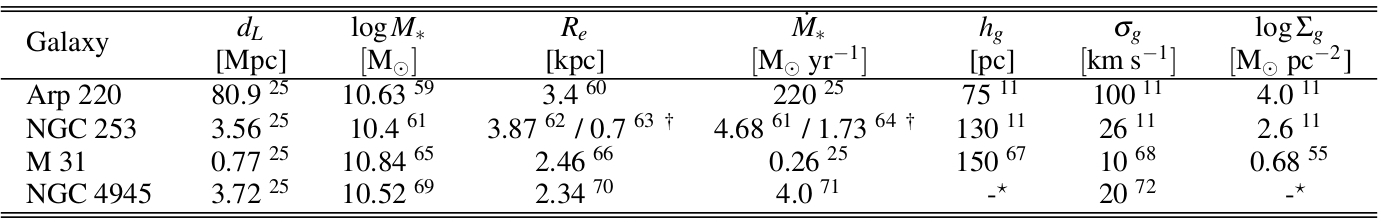}
  \caption{\textbf{Local galaxy data}
For each entry, we give a value followed by the reference from which that value is taken.
\\\hspace{\textwidth}$\star$ NGC 4945 is observed edge-on, so measurements of the gas scale height and gas surface density are unavailable. We derive them in the usual manner, as described in the Methods, using the measured gas velocity dispersion.
\\\hspace{\textwidth}$\dagger$ The gas data come exclusively from the nuclear starburst region, so we give two effective radii and SFR estimates: the first is for the entire galaxy, and the second is for the circumnuclear disk / nuclear starburst region only. We use the former for our stellar data spectrum prediction and the latter for our gas prediction.}
\end{figure*}

\renewcommand{\figurename}{Extended Data Figure}
\setcounter{figure}{0}

\begin{figure*}
  \centering
  \includegraphics[width=0.70\textwidth]{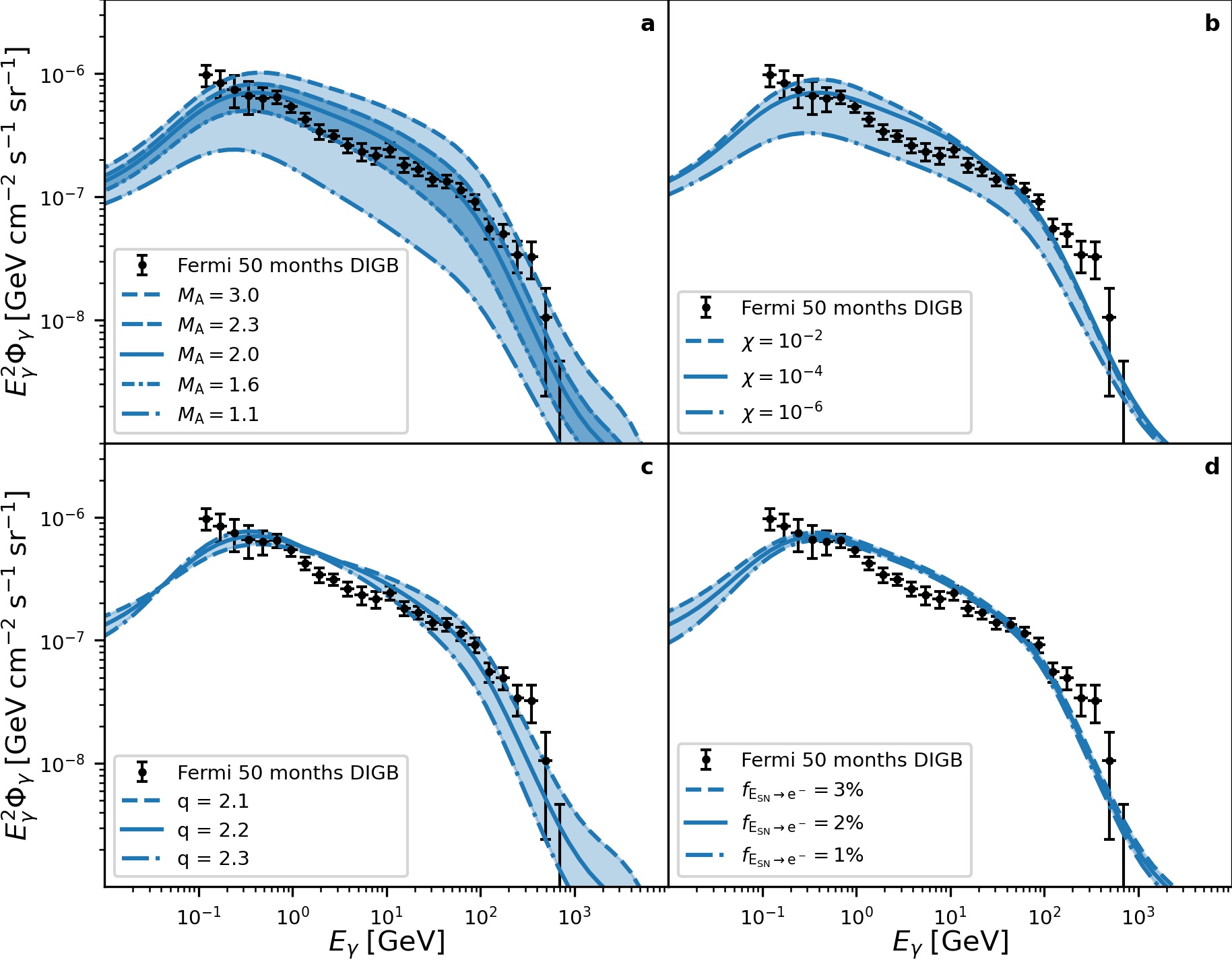}
  \caption{\textbf{The effect of varying model parameters} The plots presented here show the result of our calculations when varying the model parameters as discussed in the Supplementary Information. Our fiducial choice is plotted as a solid blue line, with the dashed and dash-dotted lines showing the spectrum for the upper and lower limits respectively of the varied parameter. The black points correspond to the \textit{Fermi} data as in Figure 4. Plot \textbf{a} shows $M_{\rm A}$ plotted for reasonable values of 1.6 and 2.3, and extremal values of 1.1 and 3.0; \textbf{b} the ionisation fraction $\chi$ for values of $10^{-2}$ and $10^{-6}$; \textbf{c} the injection index $q$ for values 2.1 and 2.3; and finally \textbf{d} the conversion fraction of supernova energy to CR electrons for values of 1\% and 3\%, which is equivalent to 10\% and 30\% of the total energy injected in all cosmic ray species. Note that varying the total CR energy budget results in a trivial scaling of the result by the same fraction, and thus is not shown.}
\end{figure*}

\begin{figure*}
  \centering
  \includegraphics[width=0.70\textwidth]{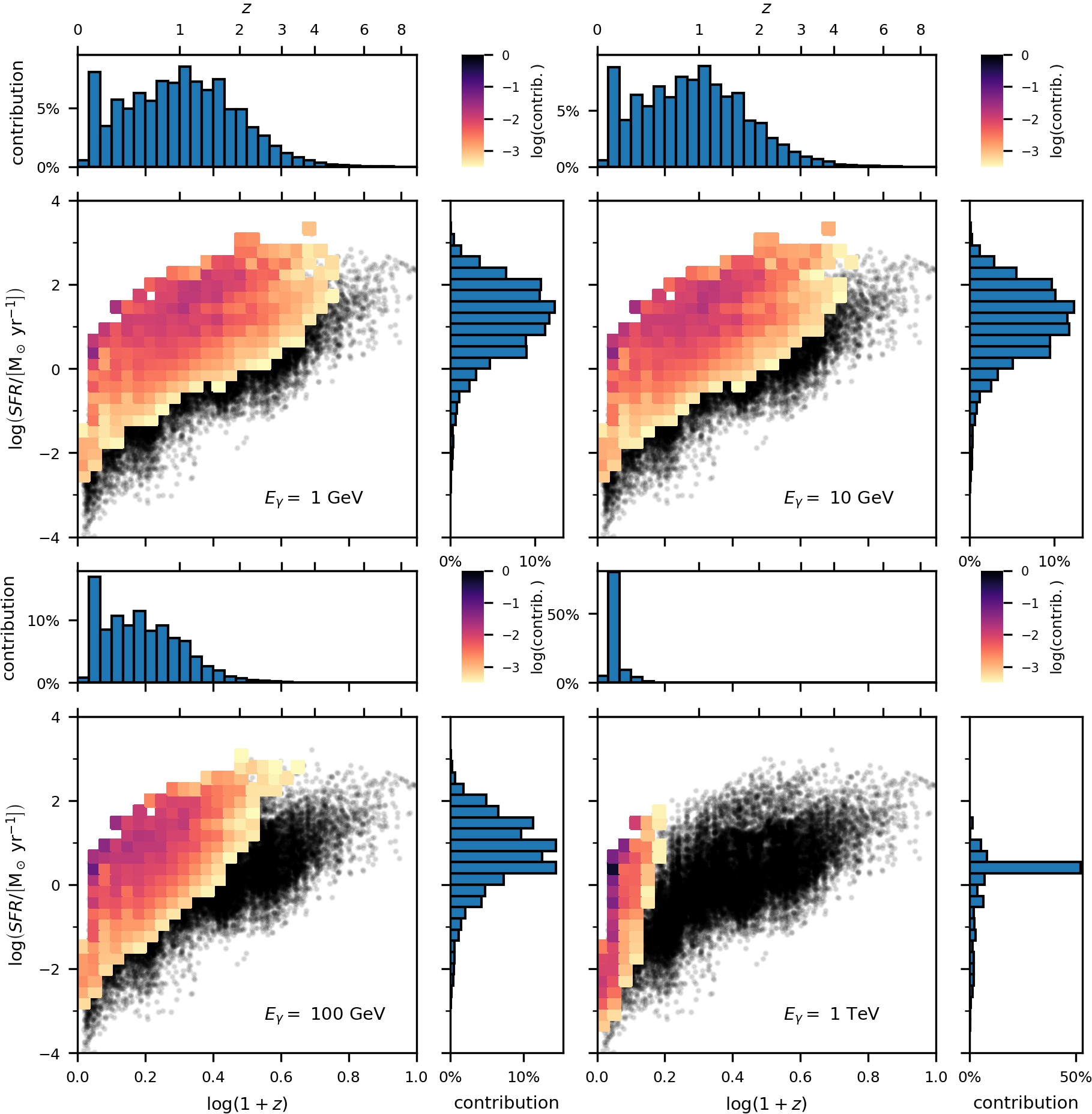}
  \caption{\textbf{The contribution of SFGs in the $\dot{M}_*$ - $z$ plane} The contribution of SFGs to the total $\gamma$-ray spectrum at selected energies in the star formation rate ($\dot{M}_*$), redshift ($z$) plane. Coloured pixels show the fractional contribution (as indicated in the colourbar) from galaxies in each bin of $\dot{M}_*$ and $z$ to the diffuse isotropic $\gamma$-ray background at the indicated energy; a fractional contribution of unity corresponds to that pixel producing all of the background, with no contribution from galaxies outside the pixel. Grey points show individual CANDELS galaxies in regions of $\dot{M}_*$ and $z$ that contribute $<10^{-3}$ of the total. Flanking histograms show the fractional contribution binned in one dimension -- $\dot{M}_*$ (right) and $z$ (top). We see that the background at lower energies is dominated by emission from galaxies on the high side of the star forming main sequence at $z\sim 1-2$, while at high energies it is dominated by the brightest systems at low redshift.}
\end{figure*}

\begin{figure*}
  \centering
  \includegraphics[width=0.70\textwidth]{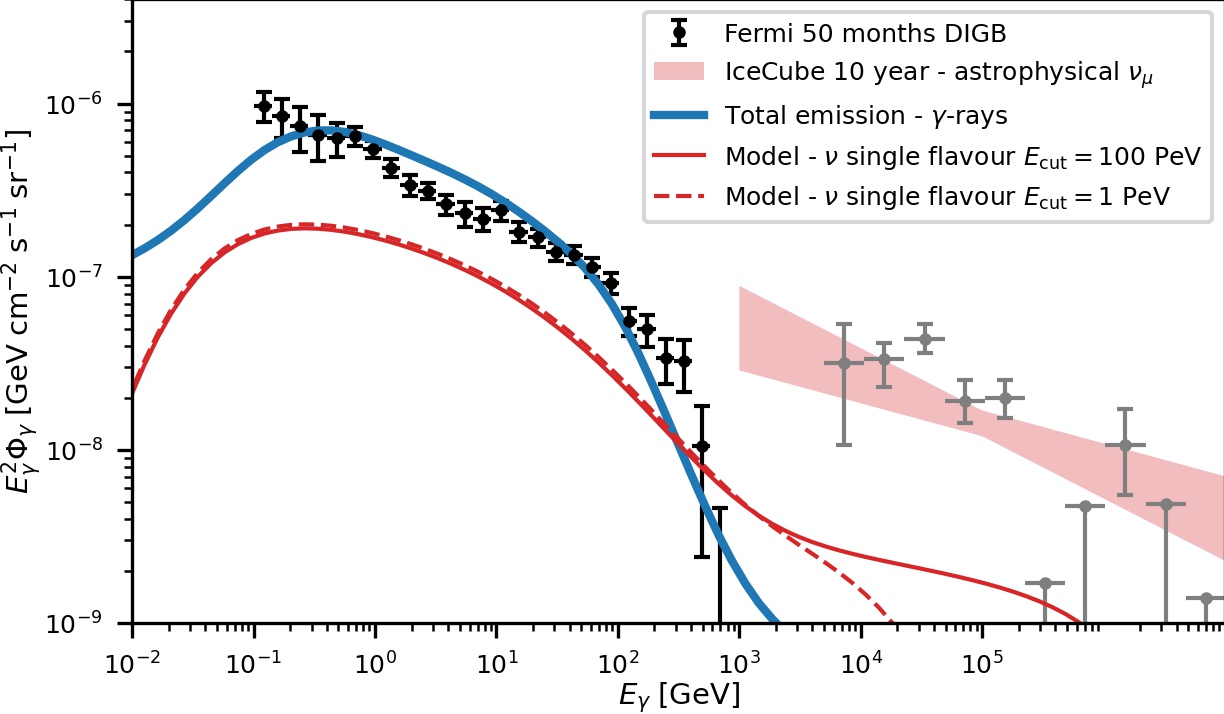}
  \caption{\textbf{The diffuse isotropic $\gamma$-ray and neutrino backgrounds} The blue line and black points show the model-predicted and observed $\gamma$-ray background, and are identical to those shown in Figure 4. The red lines show our model prediction for the neutrino background (single flavour) with $E_{\rm cut}=100$ PeV (solid line) and $E_{\rm cut}=1$ PeV (dashed line), computed as described in the Supplementary Information. We assume a neutrino flavour ratio at the detector of $\left(\nu_e : \nu_\mu : \nu_\tau\right) = \left( 1 : 1 : 1 \right)$. The red filled band shows a power law fit \textsuperscript{73} to the single flavour astrophysical neutrino background with the 90\% likelihood limit, as measured by IceCube, which is also shown as grey points, where the horizontal bars show the energy bin and the vertical bars the 1 $\sigma$ uncertainty limit.}
\end{figure*}

\begin{figure*}
  \centering
  \includegraphics[width=0.70\textwidth]{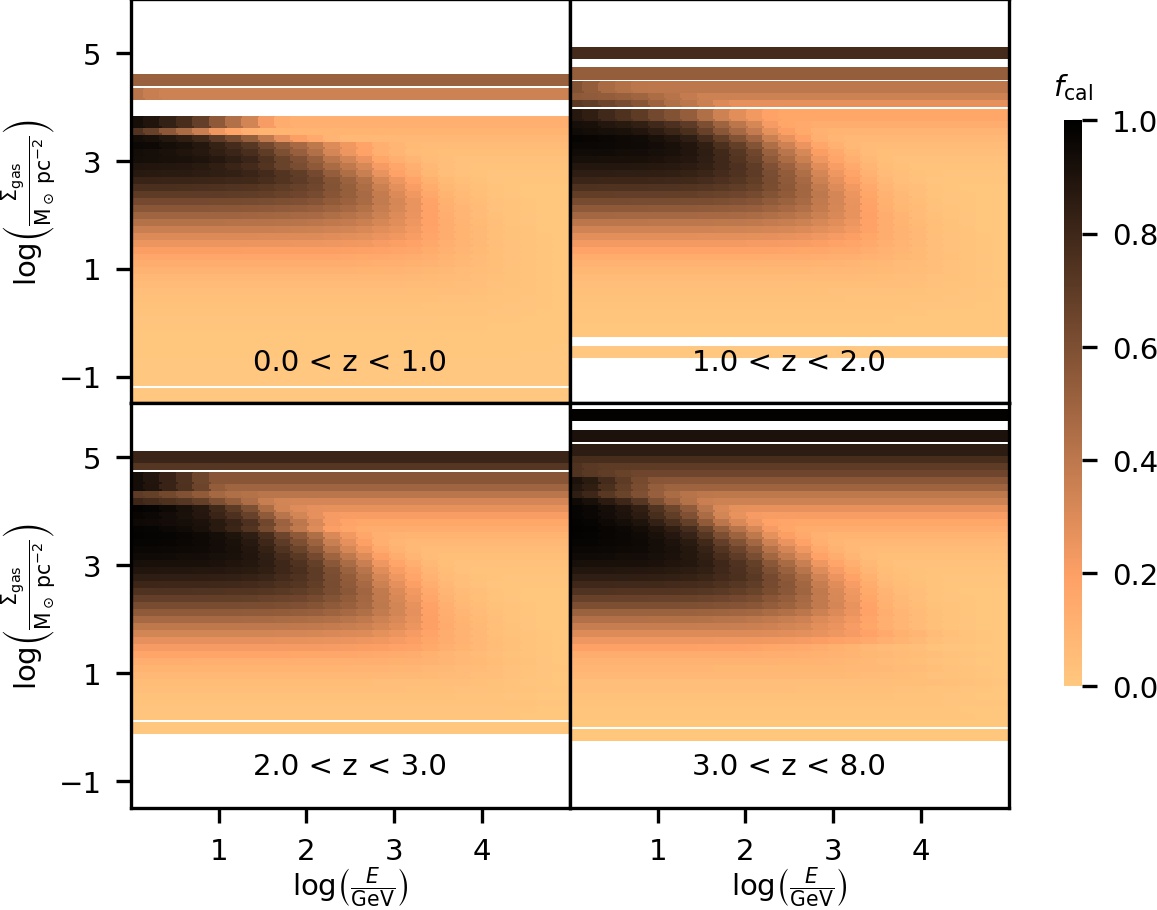}
  \caption{\textbf{Cosmic ray calorimetry in the $E$ - $\Sigma_g$ plane} Mean calorimetry fraction $f_{\rm cal}(E)$ in the surface gas density $\Sigma_g$, cosmic ray energy $E$ plane, binned in redshift intervals. This figure is constructed by deriving the gas surface density and energy dependent calorimetry fraction for each galaxy in the CANDELS sample using our model. The colour of each pixel gives the mean calorimetry fraction of all the galaxies within that particular range of $\Sigma_g$, $E$, and redshift. The horizontal white stripes correspond to ranges of $\Sigma_g$ into which no CANDELS galaxies fall for the corresponding redshift range. Several physical processes contribute to the behaviour visible in the plot. At low $\Sigma_g$, galaxies have low $f_{\rm cal}$ at all energies $E$ because there are few targets for hadronic collisions with CRs. As $\Sigma_g$ increases, the increased ISM density results in efficient calorimetry and conversion of CR energy into $\gamma$-rays for low CR energies; however, at higher energies the CR number density is low, yielding a high CR streaming velocity and rapid escape, resulting in low $f_{\rm cal}$. As $\Sigma_g$ increases further, the increasing density results in the streaming instability being suppressed efficiently by ion-neutral damping towards lower energies, reducing the calorimetry fraction further. Finally, at the highest $\Sigma_g$, the streaming instability is suppressed completely by ion-neutral damping, but streaming is still limited to the speed of light. Consequently, increasing $\Sigma_g$ further only results in increased collisions, and thus a higher calorimetry fraction.}
\end{figure*}

\begin{figure*}
  \centering
  \includegraphics[width=0.70\textwidth]{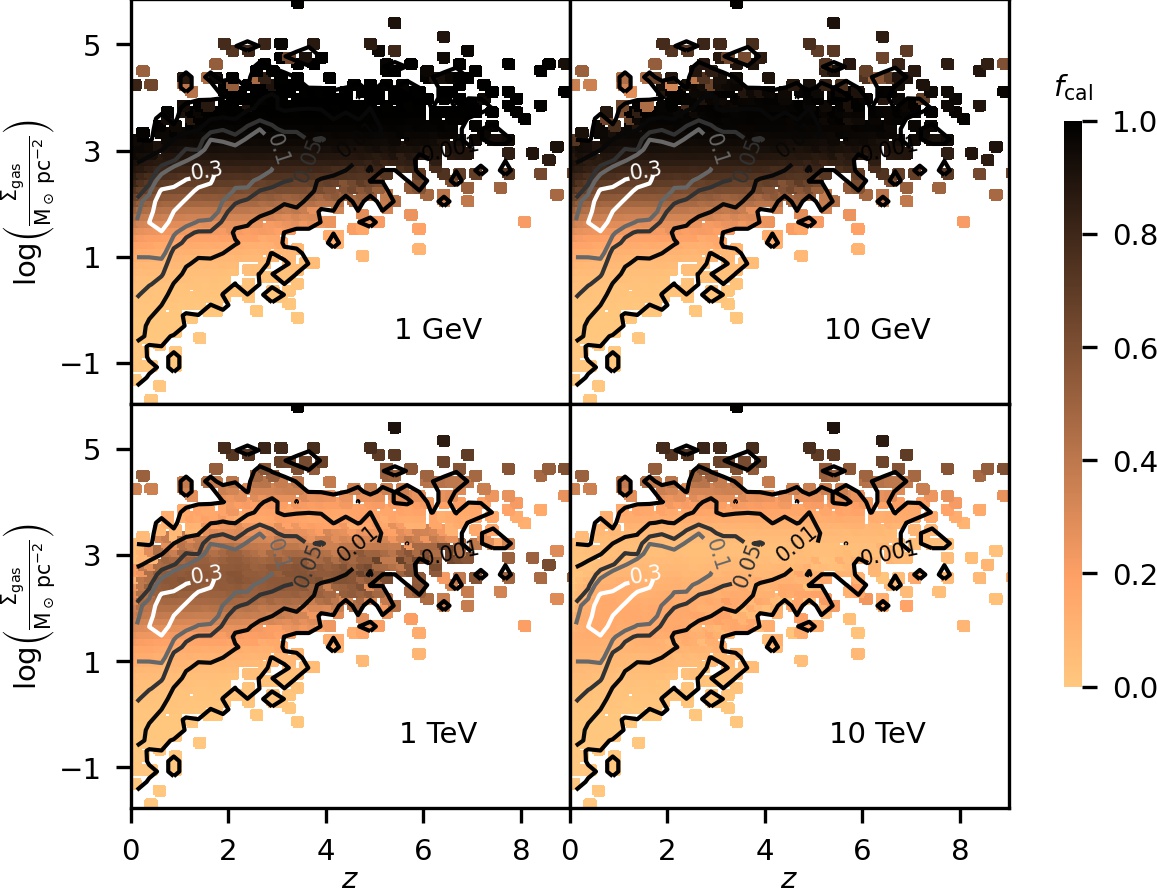}
  \caption{\textbf{Cosmic ray calorimetry in the $z$ - $\Sigma_g$ plane} Mean calorimetry fraction in the surface gas density ($\Sigma_g$), redshift ($z$) plane at CR energies $E=1$ GeV, 10 GeV, 1 TeV and 10 TeV. To construct this figure, for each CANDELS sample galaxy, we apply our model to compute $\Sigma_g$ and $f_{\rm cal}(E)$ at the indicated energies. The colour indicates the average $f_{\rm cal}(E)$ value computed over bins of ($z$, $\Sigma_g$), while contours indicate the density of the CANDELS sample in this plane. Note that the non-monotonic behaviour of $f_{\rm cal}(E)$ with $\Sigma_g$ that is most prominently visible in the 1 TeV panel is expected, for the reasons explained in the caption of Extended Data Figure 4.}
\end{figure*}

\begin{figure*}
  \centering
  \includegraphics[width=0.70\textwidth]{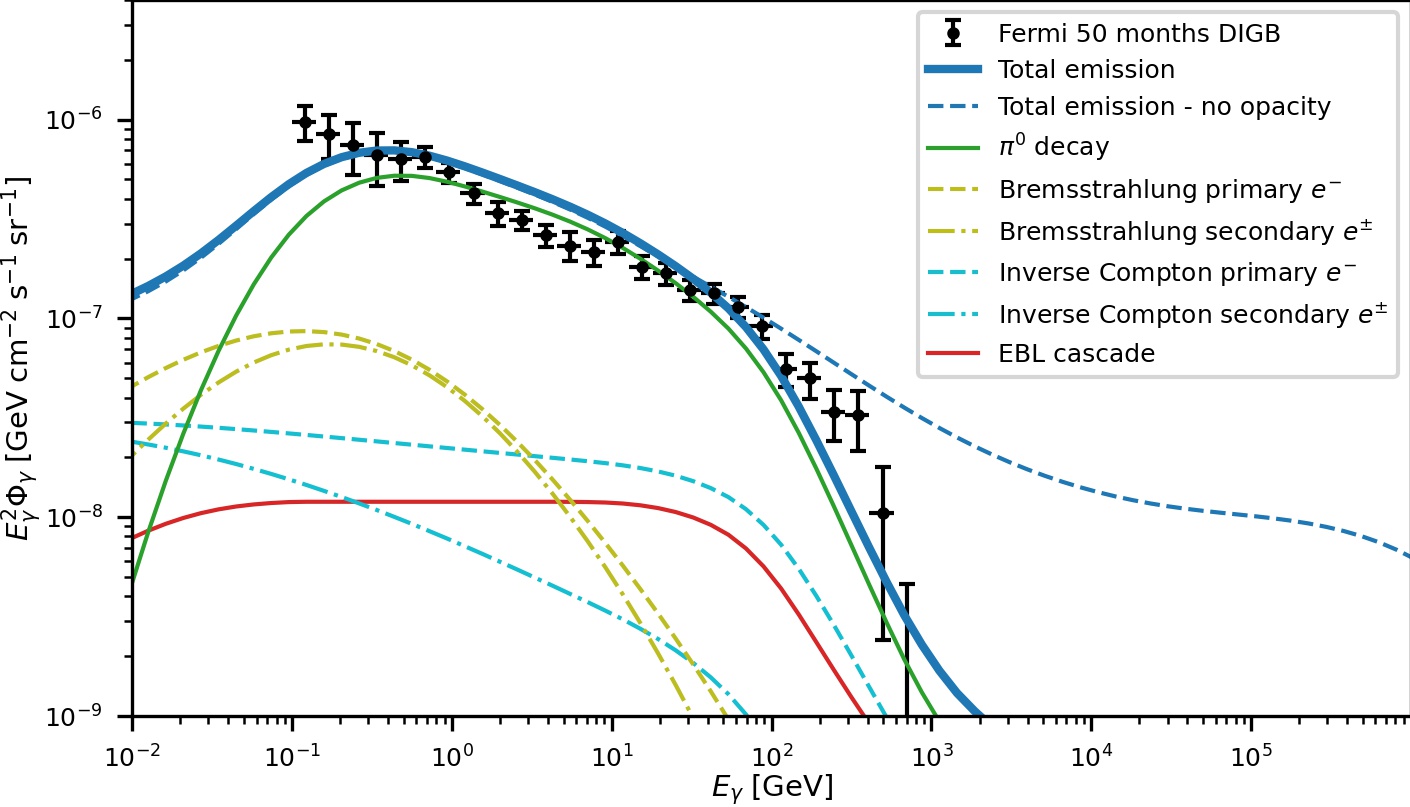}
  \caption{\textbf{Contributions to the diffuse isotropic $\gamma$-ray background} The blue line and black points show the model-predicted and observed $\gamma$-ray background, and are identical to those shown in Figure 4. The green line shows the contribution from $\pi^{0}$ decay, the olive lines the contribution from bremsstrahlung emission, and the cyan lines the contribution from the inverse Compton emission. In both cases, dashed lines show the spectrum produced by primary CR electrons and the dash-dotted lines the spectrum from secondary electrons and positrons. The red line shows the contributions from the EBL cascade.}
\end{figure*}

\newpage

\onecolumn

\setcounter{section}{0}

\counterwithout{subsection}{section}
\section*{Supplementary Information}

\subsection{Confidence intervals for source count distributions}\label{sourcecounts}
Calculation of confidence intervals on $S^2(dN/dS)$ (as shown in Figure~3) for the observed sources (both the \textit{Fermi}-observed SFGs and our model-predicted CANDELS SFGs) is non-trivial, because both surveys cover a fraction $f<1$ of the sky, and the number of sources per bin for at least some bins of $S$ is small, so we cannot compute the uncertainty by assuming that we are in the large $N$ limit. To perform the calculation, we assume that the SFG population follows a Poisson distribution on the sky (i.e., we are in the cosmological isotropic limit), so if the entire sky contains $N_{\rm tot}$ SFGs within some bin of photon flux, the probability that $N_{\rm obs}$ will be found within the observable region can be written
\begin{equation}
    P\left( N_{\rm obs} | N_{\rm tot}\right) = \frac{\left( f N_{\rm tot} \right)^{N_{\rm obs}} e^{-f N_{\rm tot}}}{N_{\rm obs}!}.
\end{equation}

We wish to solve the inverse problem, i.e., given an observed number $N_{\rm obs}$, what is the probability distribution of $N_{\rm tot}$? The answer is given by Bayes's Theorem, which requires
\begin{equation}
     P\left( N_{\rm tot} | N_{\rm obs}\right) = P\left( N_{\rm obs} | N_{\rm tot}\right) \frac{P\left( N_{\rm tot} \right)}{P\left( N_{\rm obs} \right)}
\end{equation}
where $P\left( N_{\rm tot} | N_{\rm obs}\right)$ is the posterior probability, $P\left(N_{\rm tot}\right)$ is the prior probability, and $P\left( N_{\rm obs}\right)$ is a normalisation factor. We adopt a flat prior $P\left( N_{\rm tot} \right) \propto 1$, so we can then write
\begin{equation}
     P\left( N_{\rm tot} | N_{\rm obs}\right) = \mathcal{N} \; N_{\rm tot}^{N_{\rm obs}} e^{-f N_{\rm tot}},
\end{equation}
where $\mathcal{N}$ is a normalisation constant. For $e^{-f} < 1$, which is always the case since $0<f\leq 1$, the value of $\mathcal{N}$ required to guarantee that $\sum_{N_{\rm tot}} P\left( N_{\rm tot} | N_{\rm obs}\right) = 1$ is 
\begin{equation}
    \mathcal{N} = \frac{1}{\mathrm{Li}_{-N_{\rm obs}}\left(e^{-f} \right)},
\end{equation}
where $\mathrm{Li}_s\left(z\right)$ is the polylogarithm of order $s$.

To compute the confidence interval we require the cumulative distribution function. In the discrete case this is given by calculating the probability that $N_{\rm tot} < N$, which is
\begin{eqnarray}
     P\left( N_{\rm tot} < N \right) & = & \mathcal{N} \sum^{N-1}_{n=0} n^{N_{\rm obs}} e^{-f n} 
     \\
     & = & 1 - \mathcal{N} \sum^{\infty}_{n=N} n^{N_{\rm obs}} e^{-f n}
     \\
     & = & 1 - \mathcal{N} \sum^{\infty}_{i=0} \left(i+N\right)^{N_{\rm obs}} e^{-f \left(i+N\right)}
     \\
     & = & 1 - \frac{ e^{-f N} \; \Phi\left( e^{-f}, -N_{\rm obs}, N \right)}{\mathrm{Li}_{-N_{\rm obs}}\left(e^{-f} \right)}
\end{eqnarray}
where $\Phi\left( z, s, a \right)$ is the Lerch Phi function (sometimes also referred to as the Lerch Zeta function). To obtain a particular percentile $p$ in the range 0 to 1, we simply use the continuous forms of the polylogarithm and the Lerch Phi functions, set $p = P\left( N_{\rm tot} < N \right)$ and invert the problem numerically to find the appropriate value for $N$; for the purposes of Figure~3, we are interested in the 90\% confidence interval, so we take $p=0.05$ and $p=0.95$. For the special case $N_{\rm obs} = 0$, the result simplifies to
\begin{equation}
    p = 1 - e^{-f N} \left(\frac{1}{\mathrm{Li}_0\left(e^{-f} \right)} + 1 \right),
\end{equation}
which we can invert numerically for $p=0.9$ to obtain the 90\% confidence upper limit.

In order to use the result we have just derived, we require a value for $f$. For the CANDELS data points, this is straightforward: our data come from the GOODS-S field, which has an area of $173 \; {\rm arcmin^2}$, corresponding to $f = 1.16 \times 10^{-6}$. Assigning a value of $f$ to the \textit{Fermi} data is more complex: \textit{Fermi} LAT surveys the entire sky, but it cannot detect faint sources, such as SFGs, that are too close to the Galactic plane because they are hidden by the Galactic diffuse foreground. As a result, the effective survey area depends at least somewhat on the flux and spectral shape of the target SFG -- brighter and harder sources can be detected closer to the plane than fainter and softer ones. Capturing this effect in detail would require extensive testing of the \textit{Fermi} reduction pipeline using artificial sources, which is beyond the scope of this work. For the purposes of computing the confidence intervals shown in Figure~3, we ignore this complexity, and roughly estimate that SFGs are undetectable within $15^\circ$ degrees of the Galactic plane, which corresponds to approximately $f=0.7$.

\subsection{Sensitivity of the result to model parameters}
Here we investigate the sensitivity of our results to our choice of model parameters. The set of parameters we have adopted to derive the key result of this study were deliberately chosen to be approximately the consensus value. We purposefully chose not to fine-tune our parameters to obtain a best fit, nor, more importantly, have we had to adopt extremal values of any parameter to force the calculation in the direction of making the contribution of SFGs dominant. However, it is still of interest to explore the parameter space within a reasonable range of variation. The key tunable parameters in our model are: (1) the total energy per unit mass of stars formed that is ultimately injected in CRs, (2) the Alfv\'en Mach number $M_{\rm A}$, (3) the ionisation fraction $\chi$, (4) the CR injection spectral index $q$, and (5) the ratio of CR energy injected into primary leptons to that injected into primary protons. We discuss each of these in turn.

The total CR energy budget is constrained by observations of both individual SN remnants \textsuperscript{15} and by observation of the total $\gamma$-ray luminosity of local starburst galaxies, which are generally thought to be fully calorimetric \textsuperscript{16,25,34}. These observations constrain this energy budget to be within a factor of $\approx 2$ of our fiducial choice. Varying this parameter within that range would increase or decrease our predicted background from SFGs by the same factor, while leaving the overall shape of the spectrum unchanged.

The Alfv\'en Mach number $M_{\rm A}$ cannot be measured directly in external galaxies, but is determined by the ratio of the magnetic $u_{\rm B}$ and kinetic $u_{\rm kin}$ densities in the flow, $M_{\rm A} = \left( 3 u_{\rm B}/u_{\rm kin} \right)^{-0.5}$ \textsuperscript{11}, which in turn is well constrained by dynamo theory and simulations \textsuperscript{11,74,75}. Galactic magnetic fields are driven by both turbulent and $\alpha\Omega$ dynamos. The growth timescale for the latter is the orbital period, and for the former is the eddy turnover time, which for a galaxy that is marginally stable against gravity is also comparable to the orbital period \textsuperscript{76}; since essentially all galaxies are many orbital periods old, we can assume that galactic dynamos have reached saturation. For the turbulent dynamo driven by supersonic motion, the saturation level of $u_{\rm B}/u_{\rm kin}$ is determined by the Reynolds and magnetic Prandtl numbers. 
Both of these dimensionless numbers are large in star-forming galaxies ($\mathrm{Re}\gtrsim 10^6$ \textsuperscript{76} and $\mathrm{Pr}\sim h_{\rm g}/l_{\rm AD}\gtrsim 100$, where $h_{\rm g}$ is the gas scale height and $l_{\rm AD}$ is the length scale at which magnetic fields decouple from the gas due to ambipolar diffusion \textsuperscript{11}). In this regime, Ref.~\textsuperscript{75} find $u_{\rm B}/u_{\rm kin} = 0.040 - 0.064$, which corresponds to Alfv\'en Mach numbers of 2.9 and 2.3. This provides an absolute upper limit on $M_{\rm A}$, expected from the turbulent dynamo alone. The $\alpha\Omega$ dynamo will further enhance the field strength by creating a coherent component on top of the turbulent one. Simulations in Ref.~\textsuperscript{74} find that the $\alpha\Omega$ dynamo in galactic discs saturates at $B_{\rm coh}/B_{\rm turb} = 0.27 - 0.42$, corresponding to a factor of $1.27^2 - 1.42^2$ increase in $u_{\rm B}$ compared to our estimate for the turbulent dynamo only. 
If we take the minimum value for the turbulent energy from Ref.~\textsuperscript{75} (0.04) and the minimum $\alpha \Omega$ amplification factor from Ref.~\textsuperscript{74} ($1.27^2$), this gives $M_{\rm A} = 2.27$, while the maximal values 0.064 and $1.42^2$ give $M_{\rm A} = 1.61$. Our fiducial choice of $M_{\rm A} = 2$ is simply the middle of this fairly narrow range of reasonable values. We show the results of varying $M_{\rm A}$ in Panel~\textbf{a} of Extended~Data~Figure~1, where we find that lowering $M_{\rm A}$ to 1.6 yields a 
slightly fainter background, while raising it to $M_{\rm A} = 2.3$ makes the predicted background somewhat brighter; 
the overall shape is largely unchanged in either case. For illustrative purposes we also include the extremal values of 1.1 and 3.0, corresponding to near-equipartition and very sub-equipartition field strengths.

The ionisation fraction $\chi$ has a larger plausible range: in galaxies with predominantly atomic interstellar media such as the Milky Way, it reaches $\chi\approx 10^{-2}$, while in extreme starbursts with very high densities it might reach as low as $10^{-6}$, indicating a 2 dex range around our fiducial choice \textsuperscript{11,25}. However, this parameter also enters the problem to the $1/4$ power. We show the effects of varying $\chi$ in Panel~\textbf{b} of Extended~Data~Figure~1; it is clear from this figure that, despite its larger range of variation, varying $\chi$ within its plausible range has a smaller effect than varying $M_{\rm A}$.

The injection spectral index $q$ is constrained by observations of local SN remnants, which require that it lie in the range $q\approx 2.1 - 2.3$ \textsuperscript{19,20}. We explore this range of variation in Panel~\textbf{c} of Extended~Data~Figure~1, which shows that changing the injected spectral index induces, as might be expected, a similar change in the spectral slope of the predicted background. However, given the uncertainties, it is not clear exactly which spectral index in this range would give the best agreement with the data, and we note that all of these models lie far from what would be expected for full calorimetry, as is clear from comparing Extended~Data~Figure~1 to the line for full calorimetry in Figure~4 of the main text; even for $q=2.1$ or $2.3$, the overall spectral slope is determined mostly by variation of $f_{\rm cal}(E)$, not by the choice of injection spectrum.

Finally, our fiducial choice of ratio of primary electrons to protons is motivated by what is required to reproduce the FIR-radio correlation \textsuperscript{16}. However, given various uncertainties, this could plausibly change at the factor of $\approx 2$ level. We show the effects of varying the fraction of SN energy that goes into primary CR electrons (while holding CR ions constant) in Panel~\textbf{d} of Extended~Data~Figure~1. As expected, the effects are minimal except below $\approx 1$ GeV in energy, and even there are small, since reducing the energy in primary electrons of course does not affect emission from secondaries, which are equally important.

The overall message of Extended~Data~Figure~1 is that the conclusion that SFGs contribute at least $\approx 50\%$ of the total diffuse $\gamma$-ray background seems inescapable, even if we choose extreme values for model parameters. However, there is also another, somewhat stronger conclusion to draw: within the reasonable parameter space, the spectral \textit{shape} we derive, which is dictated primarily by our model for $f_{\rm cal}(E)$, matches the spectral shape of data well over a $\gtrsim 3$ decade range in $\gamma$-ray energy. This means that, if we were to adopt extreme values of parameters such that SFGs contribute only $\approx 50\%$ of the background, leaving the remainder to be made up by some other unknown source population, matching the observed spectrum would require implausible fine-tuning: this other source population would have to produce a spectral shape and magnitude nearly identical to that of SFGs, over the entire energy range from $\approx 0.3$ GeV to $\approx 1$ TeV.

The one place where there is some minor tension between the spectral shape predicted by our model and that observed is near the EBL-induced cutoff of the spectrum between $\sim$100 GeV and $\sim$1 TeV, where our model falls slightly below the data (though the data themselves are uncertain in this energy range due to the difficulties of background subtraction). One possible explanation is that the model prediction in this energy range is sensitive to our chosen functional form for the EBL optical depth, and could conceivably be fit better by alternative models. Moreover, due to the EBL, the background at these energies is dominated by low-redshift, hard $\gamma$-ray sources (see Extended~Data~Figure~2), which are poorly sampled by the deep, narrow field of view in CANDELS. We have partly accounted for this by correcting the CANDELS star formation density using the measured star formation history, but a better solution, which we leave for future work, would be to supplement CANDELS by a wide-field, low-redshift galaxy survey. 

\subsection{Comparison to earlier work}
Our conclusion that emission from SFGs dominates the diffuse, isotropic $\gamma$-ray differs from some earlier work. It is therefore important to examine the precise reasons why this is the case. One contributing factor is that earlier models were forced to adopt single power laws for the emitted $\gamma$-ray spectrum in different classes of galaxies \textsuperscript{78,79}. By contrast, Figure~1 demonstrates that none of the four nearby resolved galaxies shown have spectra that are well described by a $\gamma$-ray spectrum in the form of a pure power law over the energy range from $E_\gamma = 1 - 1000$ GeV; our model correctly captures this behaviour, but earlier pure power law models did not. Similarly, we calculate $f_{\rm cal}$ as a function of energy directly, rather than relying on an empirical FIR-$\gamma$ correlation, and our calorimetry fractions are on average larger than those implicitly assumed in earlier works. This is because many of the lower estimates for the contribution from SFGs to the $\gamma$-ray background rely on a FIR-$\gamma$ correlation derived from early \textit{Fermi} detections of $<10$ individually-resolved SFGs \textsuperscript{7} that yields somewhat lower $\gamma$-ray luminosities than more recent fits using a larger (but still small) sample of SFGs \textsuperscript{25}, and with which our model agrees (Figure~2). Thus the reason we find that SFGs can produce the full background, whereas earlier models could not, is that our model predicts $\gamma$-ray emission that is both somewhat brighter and has a more complex spectral shape than the values adopted in earlier work.

Likewise, earlier claims that a variety of other source classes dominate the diffuse, isotropic background have also relied on extrapolated empirical correlations with large uncertainties. For instance Ref.~\textsuperscript{80} estimates the contribution from misaligned active galactic nuclei using a radio-$\gamma$ correlation derived from a sample of 16 resolved objects, coupled with a radio luminosity function extrapolated to redshifts considerably higher than those well-sampled by observations \textsuperscript{81}. By contrast, our assignment of $\gamma$-ray luminosities to SFGs is based on a physical model that agrees with local observations, and the CANDELS catalogue from which we draw our distribution of SFG properties has very good completeness over the range of redshift and star formation rate that dominates production of the diffuse, isotropic $\gamma$-ray background (see Extended~Data~Figure~2).

\subsection{Neutrinos}\label{neutrinos}
In addition to electrons and positrons, the decay of $\pi^\pm$ also produces neutrinos, which are of particular interest as they propagate largely unhindered from the source to the observer. Our goal here is to compute the all-species neutrino flux due to SFGs, so that we may compare to the astrophysical neutrino background measured by IceCube \textsuperscript{73}.

The relationship between the $\gamma$-ray and neutrino spectra is approximately given by $E_\nu^2 F_\nu \left( E_\nu = E_\gamma/2 \right) = (3/2) E_\gamma^2 F_\gamma \left( E_\gamma \right)$ \textsuperscript{82}. However, we compute the neutrino flux from the charged pion decay in our sample galaxies using the more detailed method in Ref.~\textsuperscript{35,36,37}\footnote{We caution readers, a number of recent publications calculate the neutrino spectrum using an incorrect formula for the parameterisation function $g_{\nu_{e}}(x)$ given in Ref.~\textsuperscript{35}. An Erratum has been published in Ref.~\textsuperscript{36}. Use of the incorrect formula leads to overestimation of the neutrino emission by a factor of $\sim$2.}.

Charged pion decay produces neutrinos in two steps: the initial decay of the pion creates a muon and a muon neutrino, and then the muon decays, yielding an electron, an electron neutrino, and a second muon neutrino (where we do not distinguish between particles and anti-particles). The all-flavour neutrino spectrum is then a sum over the energy distributions of all three neutrinos produced in this chain, given by
\begin{equation}
\frac{d\dot{N}_\nu}{dE_\nu}\left( E_{\rm \nu} \right) =  2 \int_{0}^{1}  \left( f_{\rm \nu_{e}}\left( x \right) + f_{\rm \nu_{\mu}^{\left(2\right)}}\left( x \right) \right) \; \frac{d\dot{N}_\pi}{dE_\pi}\left( \frac{E_{\rm \nu}}{x} \right) \; \frac{dx}{x} + \frac{2}{\lambda} \int_{0}^{\lambda}  \frac{d\dot{N}_\pi}{dE_\pi}\left( \frac{E_{\rm \nu}}{x} \right) \; \frac{dx}{x},
\label{eq:neutrino_spec}
\end{equation}
where  $\lambda = 1-\left( m_\mu / m_\pi \right)^2$, $x = E_\nu / E_\pi$, the second integral accounts for the muon neutrinos produced in the initial charged pion decay, and the functions $f_{\nu_{\rm e}}$ and $f_{\nu_{\rm \mu}^{\left(2\right)}}$ describe the energy distributions for the electron and muon neutrinos produced by decay of the secondary muon, respectively; we take them from Equations 40 and 36 of Ref.~\textsuperscript{35,36}. The ratio of neutrino flavours at the source is  $\left(\nu_e : \nu_\mu : \nu_\tau\right) = \left( 1 : 2 : 0 \right)$. However, neutrino oscillations will bring this to an even $\left(\nu_e : \nu_\mu : \nu_\tau\right) = \left( 1 : 1 : 1 \right)$ for an observer at Earth.

\autoref{eq:neutrino_spec} is the analogue to Equation 1 of the main text for $\gamma$-rays, and we can compute the resulting specific neutrino flux for each galaxy from Equation 2 of the main text simply by replacing $d\dot{N}_\gamma/dE_\gamma$ with $d\dot{N}_\nu/dE_\nu$ and setting the opacities $\tau_{\gamma\gamma} = \tau_{\rm EBL} = 0$. We use this to calculate a predicted neutrino flux from each CANDELS galaxy, and we sum to compute the neutrino background due to SFGs using Equation 3 of the main text, exactly as we do for the $\gamma$-ray background. We plot the resulting predicted neutrino spectrum in Extended~Data~Figure~3.

We find that our model predicts that SFGs produce a neutrino flux that is $\approx 15\%$ of the astrophysical neutrino background, as measured by IceCube \textsuperscript{73}, for a CR spectral cutoff energy of $E_{\rm cut}=100$ PeV. However, while the choice of $E_{\rm cut}$ has no significant effect on the $\gamma$-ray spectrum (as explained in the main text), it does matter for the neutrino spectrum due to the high energies of the astrophysical neutrinos observable by IceCube. Consequently, we find  that SFGs produce $\ll 15\%$ of the observed neutrino background if we adopt a smaller value of $E_{\rm cut}$ \textsuperscript{20}. To illustrate this, in Extended~Data~Figure~3 we show two calculations: one with our fiducial $E_{\rm cut} = 100$ PeV, and one with a smaller $E_{\rm cut} = 1$ PeV. The cutoffs in the neutrino spectrum shown in Extended~Data~Figure~3 are a direct result of the adopted value of $E_{\rm cut}$. We also note that the normalisation of our predicted neutrino spectrum is sensitive to bright, hard neutrino sources at low redshift, which dominate at high energy but are poorly sampled by the small CANDELS field of view. This suggests that it would be worthwhile in the future to repeat this analysis using a survey of SFGs that is wider but shallower than CANDELS. A further consideration relates to the exact form of the cosmic ray calorimetry at ultra high energies. Here we would expect cosmic rays with sufficiently large gyro radii to interact with and scatter off the external turbulence above the damping scale, significantly decreasing the diffusion coefficients and increasing calorimetry in turn. We leave an exploration of this mechanism and the injection cutoff energy for future work.

\section*{Supplementary Information References}

\begin{enumerate}
\setcounter{enumi}{73}
\item Bendre, A., Gressel, O. \& Elstner, D. Dynamo saturation in direct simulations of the multi-phase turbulent interstellar medium. \textit{Astron. Nachr.} \textbf{336}, 991. doi:10.1002/asna.201512211 (2015).
\item Federrath, C. Magnetic field amplification in turbulent astrophysical plasmas. \textit{J. Plas. Phys.} \textbf{82}, 535820601. doi:10.1017/S0022377816001069 (2016).
\item Krumholz, M. R., Burkhart, B., Forbes, J. C. \& Crocker, R. M. A unified model for galactic discs: star formation, turbulence driving, and mass transport. \textit{Mon. Not. R. Astron. Soc.} \textbf{477}, 2716–2740. doi:10.1093/mnras/sty852 (2018).
\item Krumholz, M. R. The big problems in star formation: The star formation rate, stellar clustering, and the initial mass function. \textit{Phys. Rep.} \textbf{539}, 49–134. doi:10.1016/j.physrep.2014.02.001 (2014).
\item Ando, S. \& Pavlidou, V. Imprint of galaxy clustering in the cosmic gamma-ray background. \textit{Mon. Not. R. Astron. Soc.} \textbf{400}, 2122–2127. doi:10.1111/j.1365-2966.2009.15605.x (2009).
\item Tamborra, I., Ando, S. \& Murase, K. Star-forming galaxies as the origin of diffuse high-energy backgrounds: gamma-ray and neutrino connections, and implications for starburst history. \textit{J. Cosmology Astropart. Phys.} \textbf{2014}, 043. doi:10.1088/1475-7516/2014/09/043 (2014).
\item Hooper, D., Linden, T. \& Lopez, A. Radio galaxies dominate the high-energy diffuse gamma-ray background. \textit{J. Cosmology Astropart. Phys.} \textbf{2016}, 019. doi:10.1088/1475-7516/2016/08/019 (2016).
\item Willott, C. J., Rawlings, S., Blundell, K. M., Lacy, M. \& Eales, S. A. The radio luminosity function from the low-frequency 3CRR, 6CE and 7CRS complete samples. \textit{Mon. Not. R. Astron. Soc.} \textbf{322}, 536–552. doi:10.1046/j.1365-8711.2001.04101.x (2001).
\item Anchordoqui, L. A., Goldberg, H., Halzen, F. \& Weiler, T. J. Neutrino bursts from Fanaroff Riley I radio galaxies. \textit{Phys. Let. B} \textbf{600}, 202–207. doi:10.1016/j.physletb.2004.09.005 (2004).
\end{enumerate}

\end{document}